\documentclass[sigconf]{aamas} 

\usepackage{balance} 
\usepackage{subfigure}
\usepackage{multirow}
\usepackage{pifont}
\usepackage{multicol}
\usepackage{graphicx}
\usepackage{algorithmic}
\usepackage[linesnumbered,ruled,vlined]{algorithm2e}

\theoremstyle{thmstyletwo}%

\theoremstyle{thmstylethree}%
\newtheorem{definition}{Definition}%

\setcopyright{ifaamas}
\acmConference[AAMAS '24]{Proc.\@ of the 23rd International Conference
on Autonomous Agents and Multiagent Systems (AAMAS 2024)}{May 6 -- 10, 2024}
{Auckland, New Zealand}{N.~Alechina, V.~Dignum, M.~Dastani, J.S.~Sichman (eds.)}
\copyrightyear{2024}
\acmYear{2024}
\acmDOI{}
\acmPrice{}
\acmISBN{}

\acmSubmissionID{<<319>>}

\title[AAMAS-2024 Formatting Instructions]{Applying Opponent Modeling for Automatic Bidding in Online Repeated Auctions}

\author{Yudong Hu}
\affiliation{
  \institution{University of Chinese Academy of Sciences}
  \city{Beijing}
  \country{China}}
\email{huyudong201@mails.ucas.ac.cn}

\author{Congying Han}
\authornote{Corresponding author.}
\affiliation{
  \institution{University of Chinese Academy of Sciences}
  \city{Beijing}
  \country{China}}
\email{hancy@ucas.ac.cn}

\author{Tiande Guo}
\affiliation{
  \institution{University of Chinese Academy of Sciences}
  \city{Beijing}
  \country{China}}
\email{tdguo@uacs.ac.cn}

\author{Hao Xiao}
\affiliation{
  \institution{Institute of Electrical Engineering, Chinese Academy of Sciences}
  \city{Beijing}
  \country{China}}
\email{xiaohao09@mail.iee.ac.cn}

\begin{abstract}
Online auction scenarios, such as bidding searches on advertising platforms, often require bidders to participate repeatedly in auctions for identical or similar items. Most previous studies have only considered the process by which the seller learns the prior-dependent optimal mechanism in a repeated auction. However, in this paper, we define a multiagent reinforcement learning environment in which strategic bidders and the seller learn their strategies simultaneously and design an automatic bidding algorithm that updates the strategy of bidders through online interactions. We propose Bid Net to replace the linear shading function as a representation of the strategic bidders' strategy, which effectively improves the utility of strategy learned by bidders. We apply and revise the opponent modeling methods to design the PG (pseudo-gradient) algorithm, which allows bidders to learn optimal bidding strategies with predictions of the other agents' strategy transition. We prove that when a bidder uses the PG algorithm, it can learn the best response to static opponents. When all bidders adopt the PG algorithm, the system will converge to the equilibrium of the game induced by the auction. In experiments with diverse environmental settings and varying opponent strategies, the PG algorithm maximizes the utility of bidders. We hope that this article will inspire research on automatic bidding strategies for strategic bidders.
\end{abstract}

\keywords{Auction Theory; Strategic Bidder; Opponent Modeling; Multiagent Reinforcement Learning}

\newcommand{\BibTeX}{\rm B\kern-.05em{\sc i\kern-.025em b}\kern-.08em\TeX}

\makeatletter
\gdef\@copyrightpermission{
	\begin{minipage}{0.3\columnwidth}
		\href{https://creativecommons.org/licenses/by/4.0/}{\includegraphics[width=0.90\textwidth]{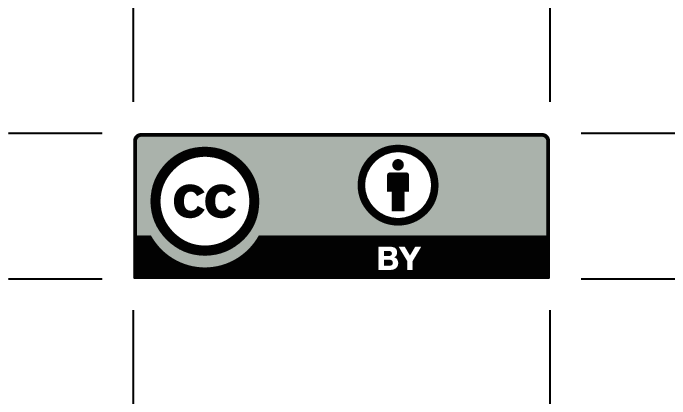}}
	\end{minipage}\hfill
	\begin{minipage}{0.7\columnwidth}
		\href{https://creativecommons.org/licenses/by/4.0/}{This work is licensed under a Creative Commons Attribution International 4.0 License.}
	\end{minipage}
	\vspace{5pt}
}
\makeatother

\begin{document}


\pagestyle{fancy}
\fancyhead{}


\maketitle 


\section{Introduction}\label{sec1}

The rapid development of electronic commerce has generated a significant demand for online auctions \cite{no41}. Employing appropriate mechanisms to facilitate the sale of items, such as advertisements on electronic platforms, can effectively improve revenue \cite{no1,no2,no27,no28}. Among the most prevalent mechanisms in online auctions are the VCG auction (Vickrey-Clarke-Groves) \cite{no4,no5} and the GSP auction (Generalized Second Price) \cite{no1,no3}. Both of these auctions are instances of the prior-independent auction mechanism, which can be executed efficiently, but does not maximize revenue.

With the advent of deep reinforcement learning, it has become feasible to learn optimal auctions and maximize seller revenue \cite{no35,no51}. Deep neural network techniques, such as Myerson Net \cite{no8}, ALGNet \cite{no9}, MenuNet \cite{no34}, and Deep GSP \cite{no36}, have been applied to derive optimal mechanisms based on bidding data, demonstrating efficiency even in multi-bidder, multi-item auction scenarios. These methods incorporate prior-dependence of bidders' value distribution to the mechanism through the utilization of bidding data sourced from online auctions. By creating mechanisms that adhere to the principle of Incentive Compatibility (IC), the seller can collect truthful bidding data from online auctions and employ them to estimate the value distribution among bidders \cite{no29}.

Unlike the seller's perspective, recent studies have introduced strategic bidding approaches in online auctions \cite{no52}. These methods exploit the process in which the seller learns the value distribution of the bidders using bid samples \cite{no30,no31} and design strategic bidders to increase the utility with specific bidding strategies. The distribution reporting model \cite{no10} states that in incentive compatibility auctions, bidders can improve their utility by bidding according to a specific "fake distribution". When the seller employs the Myerson Net learning mechanism, adversarial learning strategies maximize the utility of strategic bidders \cite{no11}, particularly when other participants in the system adopt truthful bidding strategies \cite{no12}.

The primary aim of this paper is to formulate an approach that enables bidders to automatically learn bidding strategies. Our approach diverges from previous research on strategic bidding, such as ZIP (Zero Intelligence Plus) \cite{no39,no40}, which focused on the learning process of bidders through repeated interactions within static auction mechanisms. In our work, we depict the repeated auction environment as a multiagent reinforcement learning system, where both the seller and individual bidders are treated as agents. Each agent employs independent algorithms to acquire strategies. Although bidders engage in repeated auctions for the same item, it is important to note that the value distribution and strategies of agents can evolve over time. This requires bidders to predict changes in the mechanism induced by bidding strategies and learn how to bid in response to a dynamic mechanism.

Opponent modeling research offers robust approaches to strategy learning within the aforementioned system, where all agents independently update their strategies with strategy prediction \cite{no53}. By anticipating transitions in the opponent's strategy, an agent can acquire strategies that yield superior rewards in repeated games. In particular, LOLA (Learning with Opponent-Learning Awareness) \cite{no18} is the first algorithm capable of achieving cooperation in the Prisoner's Dilemma game through independent opponent modeling processes for each agent. Subsequent research, COLA (Consistency in Opponent-Learning Awareness) \cite{no20}, states that the system will reach the target equilibrium when the predictions of the opponent's strategy (lookahead rate) maintain consistency.

In this paper, we provide a complete definition of automatic bidding for bidders in online auctions and design algorithms with theoretical support. Our contributions are summarized as follows: 1) We model the behavior of the seller and bidders in repeated auctions as an induced game in which all agents learn strategies in repeated interactions. We adopt Myerson Net as the seller strategy and design an automatic bidding strategy for bidders. Our objective is that this MARL system converges to an equilibrium in which bidders have optimal utility. 2) We propose Bid Net to represent the strategies learned by the bidders. The network structure of the Bid Net has the advantage of satisfying the IR (individual rationality) and accurate gradient propagation with NeuralSort. We illustrate that the Bid Net is an efficient improvement of the simple linear strategy through experiment. 3) We design the PG (pseudo-gradient) algorithm based on the opponent modeling method. The PG algorithm is an automatic bidding method for updating the bidding strategy based on the prediction of changes in the parameters of the auction mechanism. We prove that when a bidder employs the PG algorithm, it can learn the optimal bidding strategy under the current prior-dependent mechanism. When all bidders adopt the PG algorithm, the system will converge to the Nash equilibrium of the induced game.

In our experiments, we compare the Bid Net with the previously used linear shading bidding strategy \cite{no11}. The results show that Bid Net can learn utility-maximizing bidding strategies, even in environments where the parameters of the seller's mechanism are constantly updated through learning. In an experiment in which all bidders use the same algorithm learning strategy, the results show that only the PG algorithm can stably learn the target equilibrium strategy of the induced game and maximize the average utility of the bidders. To illustrate the effectiveness of our automatic bidding algorithm, we test it in different environmental settings and with various opponent strategies. The PG algorithm achieves higher utility when other bidders use a static or dynamic strategy and can be applied to arbitrary environmental parameter settings. These experiments illustrate the applicability of our automatic bidding algorithm to online repeated auctions in terms of effectiveness and generalizability.


\section{Notation and Background}

In this paper, we focus on single-item auctions as defined in \cite{no49}. The values of the $n$ bidders are drawn from their value distributions $v_i \sim F_i$. $F = F_1 \times \dots \times F_n$ is the joint distribution. For each bidder $i$, its bidding function is denoted as $B_i$, with the actual bid being $b_i = B_i(v_i)$. We use $\pi_i$ to denote the strategy of bidder $i$, when it comes from the learnable parameters. An example of a bidding strategy is the linear shading strategy, expressed as $B_i(v_i) = \alpha_i \cdot v_i$. The truthful bidding strategy $B_i(v_i)=v_i$ is the simplest strategy.

We use $M$ to represent the seller's mechanism, which comprises both the allocation rule $\vec{a}$ and the payment rule $\vec{p}$. The seller receives joint bids $\vec{b}=(b_1,\cdots,b_n)$ and generates allocation $\vec{a}(\vec{b}) = (a_1(\vec{b}),\cdots,a_n(\vec{b}))$ and payment $\vec{p}(\vec{b}) = (p_1(\vec{b}),\cdots,p_n(\vec{b}))$ according to its mechanism $(\vec{a},\vec{p})=M(\vec{b},\theta)$. Here, $\theta$ signifies the strategy parameter for the seller. In the context of the traditional first-price auction, the item is allocated to the bidder with the highest bid, denoted as $(a_i = 1 \ \text{if} \ b_i=\max(\vec{b}) \ , \text{otherwise} \ 0)$, and the payment equals the bid of the winning bidder, indicated as $(p_i=b_i \ \text{if} \ a_i=1 \ , \text{otherwise} \ 0)$.

\subsection{Traditional Prior-dependent Auction}

The main results of auction theory have been consolidated in \cite{no37}. In addition, \cite{no38} discussed the representation and reasoning processes associated with auctions. For prior-independent mechanisms, such as fisrt-price and second-price auctions, the seller's revenue is the same when the bidders respond optimally \cite{no7}. This has inspired research on prior-dependent auctions with IC constraint, where the truthful bidding strategy is the dominant strategy for bidders. 

The well-known revenue-maximizing mechanism is the Myerson auction \cite{no7}, which is the optimal mechanism that satisfies the IC constraint under the premise that the value of the bidders is public information. In the Myerson auction, each bidder's value $v_i$ is transformed into a virtual value $w_i = g_i(v_i)$, after which a second-price auction with a reserve price of 0 is conducted using the set of virtual values $\vec{w}$. The virtual value function $g_i$ is determined by the distribution $F_i$ and the density function $f_i$ associated with $v_i$.
\begin{equation*}
g_i(v_i) = v_i - \frac{1-F_i(v_i)}{f_i(v_i)}.
\end{equation*}
In fact, the virtual value function $g_i$ plays a key role in determining both the allocation and payment of the mechanism. 

At this point, it is assumed that the joint value distribution of the other bidders is represented as $F_{-i}$. Consequently, the utility of the bidder $i$ can be expressed as:
\begin{equation*}
U_i(\vec{a},\vec{p},v_i) = \int_{F_{-i}} [a_i(b_i,B_{-i}(v_{-i}))v_i-p_i(b_i,B_{-i}(v_{-i}))] f_{-i}(v_{-i}) dv_{-i}.
\end{equation*}
Here, $dv_{-i} = dv_1dv_2\cdots dv_{i-1}dv_{i+1}\cdots dv_n$ and $b_i=B_i(v_i)$. In the context of single-item auctions, it is necessary for the allocation to be 0 or 1. As a result, the above equation can be reformulated as:
\begin{equation*}
U_i(\vec{a},\vec{p},v_i) = \int_{F_{-i}} \mathbb{I}(a_i=1)     [v_i-p_i(b_i,B_{-i}(v_{-i}))] f_{-i}(v_{-i}) dv_{-i},
\end{equation*}
Here, $\mathbb{I}$ denotes the indicator function, and $a_i=a_i(b_i,B_{-i}(v_{-i}))$. The seller's revenue can be defined as:
\begin{equation*}
R(\vec{a},\vec{p},F) = \int_{F} \sum_{i=1 \cdots n}  p_i(\vec{b}) \ dv_{1} \cdots dv_{n}.
\end{equation*}
The Myerson mechanism has been proven to satisfy the IC constraint, indicating that the truthful bidding strategy $B_i(v_i)=v_i$ is the optimal response to the Myerson mechanism \cite{no7}.

We consider the standard setting as an example. There are two bidders and their value distribution is identical, represented as $v_i \sim U[0,1]$. If the seller employs the Myerson mechanism, the virtual value function is $g_i(v_i) = 2v_i-1$. In response, both bidders choose to adopt the truthful bidding strategy as their best response to the Myerson mechanism. The expected utility of each bidder is $\frac{1}{12}$ and the expected revenue of the seller is $\frac{5}{12}$. However, if the seller adopts the first-price or second-price auction instead, the revenue generated is only $\frac{1}{3}$.

\begin{figure}[]
  \centering
  \includegraphics[width=0.9\linewidth]{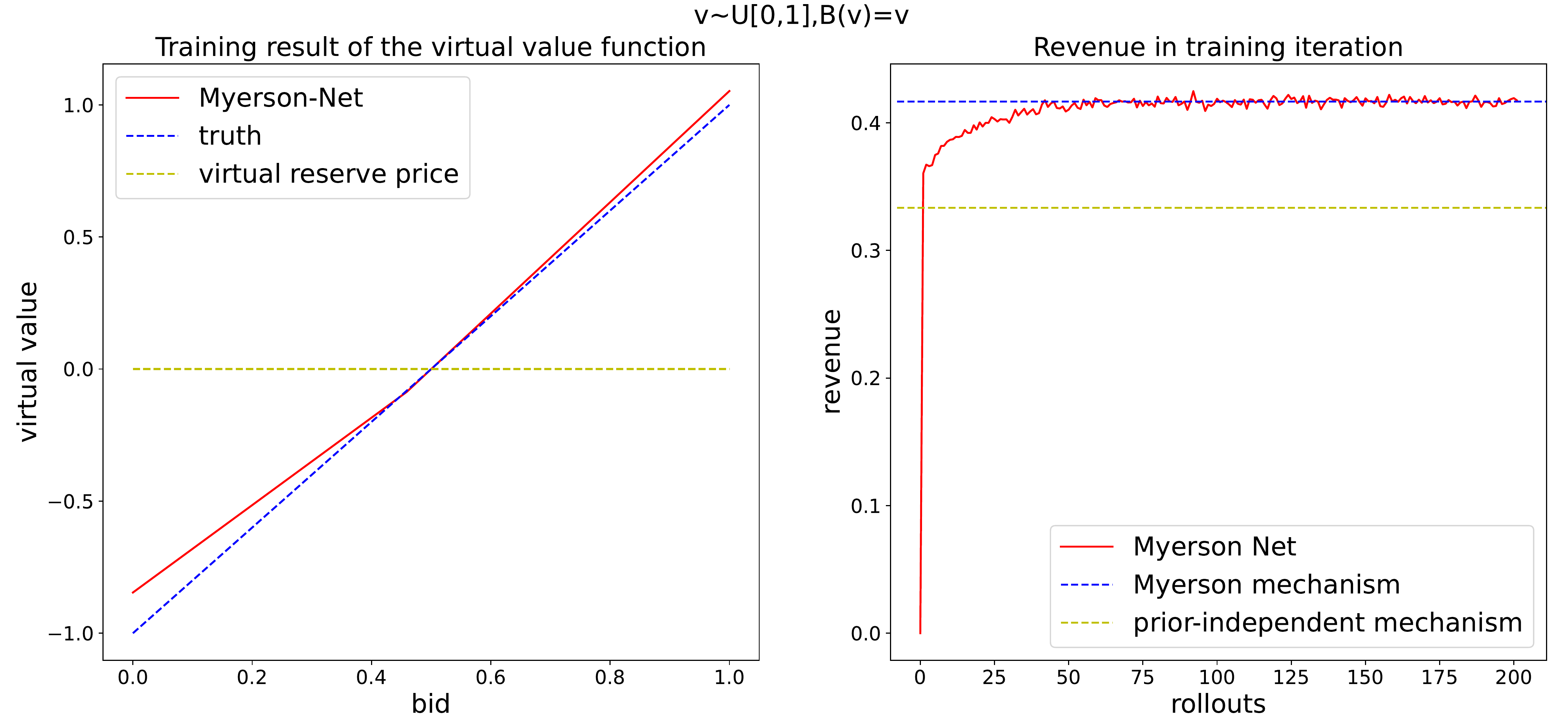}
  \vspace{-0.2cm}
  \caption{The learned virtual value function after 200 iterations of Myerson-Net in single-item, two-bidder auction. The line labeled truth indicates that it is derived from Myerson's Lemma, and the line labeled Myerson-Net is from the output of the network. The revenue curve represents the revenue of the mechanism in the learning process.}
  \label{fig:vv}
  \Description{virtual value.}
  \vspace{-0.4cm}
\end{figure}

\subsection{Online Repeated Auction and the Myerson Net}

In this paper, online auction means the auctions held in online marketplaces, such as bidding searches on advertising platforms \cite{no50}. In this context, bidders frequently participate in multiple auctions for identical or similar items. This circumstance presents opportunities to implement prior-dependent mechanisms, even when access to the value distribution of bidders is limited. The seller can establish a mechanism that adheres to the IC constraints, requiring bidders to employ a truthful bidding strategy. By collecting bid samples from repeated auctions, the seller can estimate the value distribution of the bidders, denoted as $\vec{F}$, and subsequently adjust the mechanism parameters to converge towards the optimal Myerson auction.

Myerson Net \cite{no8} presents a technique that leverages a monotonely increasing parameterized neural network $G$ to acquire virtual value functions $\vec{g}$ and derive optimal auction mechanisms.
\begin{definition}[Myerson Net] 
For joint bids $\vec{b}=(b_1,b_2,\cdots,b_n)$ from bidders $i=1\cdots n$, the seller first transforms each bid into a virtual value $w_i = G_i(b_i,\theta)$ ($\theta$ is the network parameter). The item is then assigned to the bidder with the highest positive virtual value. Payment is the minimum bid required for the winner to win:
\begin{equation*}
p_i = \mathbb{I} (a_i = 1) b^* _i, \quad b^* _i = \mathop{\text{argmin}}\limits_{b_i} [G_i(b_i,\theta) = \mathop{\text{max}}\limits_{j \in 1\cdots n}  G_j(b_j,\theta) > 0].
\end{equation*}
\end{definition}
From the definition we can obtain the following property.
\begin{theorem}
Myerson Net satisfies the IC constraints in single-item auctions. For a Myerson Net $M$, we add a strictly increasing function $Q$ to each virtual value function $G'_i=Q \circ G_i$ that satisfies $G_i(b_i) = 0 \iff Q \circ G_i(b_i) = 0$ to obtain another mechanism $M'$. Then $M'$ and $M$ have the same allocation and payment rule.
\end{theorem}

The proof of Theorem 2.1 is given in the supplementary metarial. We use $\cong$ to represent the two different networks with the same allocation and payment mechanism. Since the Myerson Net satisfies the IC constraints, we can assume that the bidders will adopt the truthful bidding strategy. Using the real bids obtained in repeated auctions, the seller updates the network parameter $\theta$ with the objective of maximizing revenue $r$, which results in the convergence of the virtual value function $G$ to the optimal function $\vec{g}$ corresponding to the true value distribution of the bidders. This suggests that the system converges to the revenue-maximizing Myerson mechanism when bidders bid truthfully in this environment.

Figure 1 provides an example of the application in the standard setting. The figure on the left shows that the seller has learned the optimal virtual value function in 200 iterations. The figure on the right shows the change in revenue in the learning iterations. We can see that the seller using Myerson Net obtains optimal revenue, which exceeds the prior-independent mechanism.

\subsection{Strategic Bidder and the Induced Game}

Myerson Net provides an approach as a revenue-maximizing mechanism for online repeated auctions. Since Myerson Net satisfies the incentive compatibility constraint, the optimal strategy for bidders in a single auction round is the truthful bidding strategy. However, some previous studies have found that bidders in repeated auctions have more efficient strategies, which requires them to bid according to some specific distribution instead of true value distribution. Although bidders initially lose some utility under the IC-constrained mechanism, untruthful bids can lead the seller to misestimate the value distribution and increase bidders' long-term utility. 

Tang's research \cite{no10} defines the induced game of auction when the seller adjusts the mechanism according to the bids. As an example, the Myerson mechanism induces a game in which the players contain only bidders. In this game, bidders adopt specific bidding strategies rather than the truthful strategy, and their utility is derived from the Myerson mechanism that runs based on bids.

\begin{definition}[Induced game of Myerson mechanism $M$]
The induced game is represented as $(N,A,\vec{U})$, where $N=\{1,\cdots,n\}$ is the set of bidders, $A=\{B_1,\cdots,B_n\}$ is the bidding function set, and $\vec{U}=(U_1,\cdots,U_n)$ is the utility function. Given the joint action $(B_1,\cdots,B_n)$, the utility is derived by applying the Myerson mechanism $M$ with the assumption $B_i(v_i)=v_i$.
\end{definition}
We present the partial payment matrix for the induced game of the single-item, two-bidder auction in the supplementary material and we have the following property.
\begin{theorem}
We assume that there are two bidders and their value distribution is $U[0,1]$ (standard setting), the induced game is as detailed in Definition 2. For strategic bidders with a limited strategy space $B(v) = \alpha v$ (referred to as the linear shading strategy space), the Nash equilibrium of the induced game yields $B_i(v_i) = \frac{5}{14} v_i$. However, in cases where bidders have access to arbitrary monotone increasing strategies, a Nash equilibrium emerges with bidding strategies represented as $B_i(v_i) = \frac{v_i+1}{4}$.
\end{theorem}
For the two bidders in the standard setting, the expected utility in each round of the auction when they employ a truthful bidding strategy is $\frac{1}{12}$. When their bidding strategy is $B_i(v_i) = \frac{v_i+1}{4}$, the expected utility is $\frac{1}{6}$. This suggests that strategic bidders can increase utility through specific bidding strategies in the induced game. The proof of Theorem 2.2 is given in the supplementary material.


\section{Muitiagent Reinforcement Learning Based Automatic Bidding Method}

To analyze the equilibrium and learning process of the induced game in Section 2.3, we introduce the environment where the seller and strategic bidders learn strategies in a repeated auction for maximum reward. We assume that the value distribution of the bidders is unavailable to the seller. Bidders and the seller will update their strategies based on the results of each auction round and the corresponding rewards. The above single-item auction is repeated until the strategies of all agents converge.

\subsection{Bid Net}

From Theorem 2.2 we can see that bidders learn different equilibria when their strategy space is limited. Since an optimal mechanism can be learned through strategy networks, we consider setting up a similar structure for the strategy learning of bidders.

\begin{figure}[]
  \centering
  \includegraphics[width=0.8\linewidth]{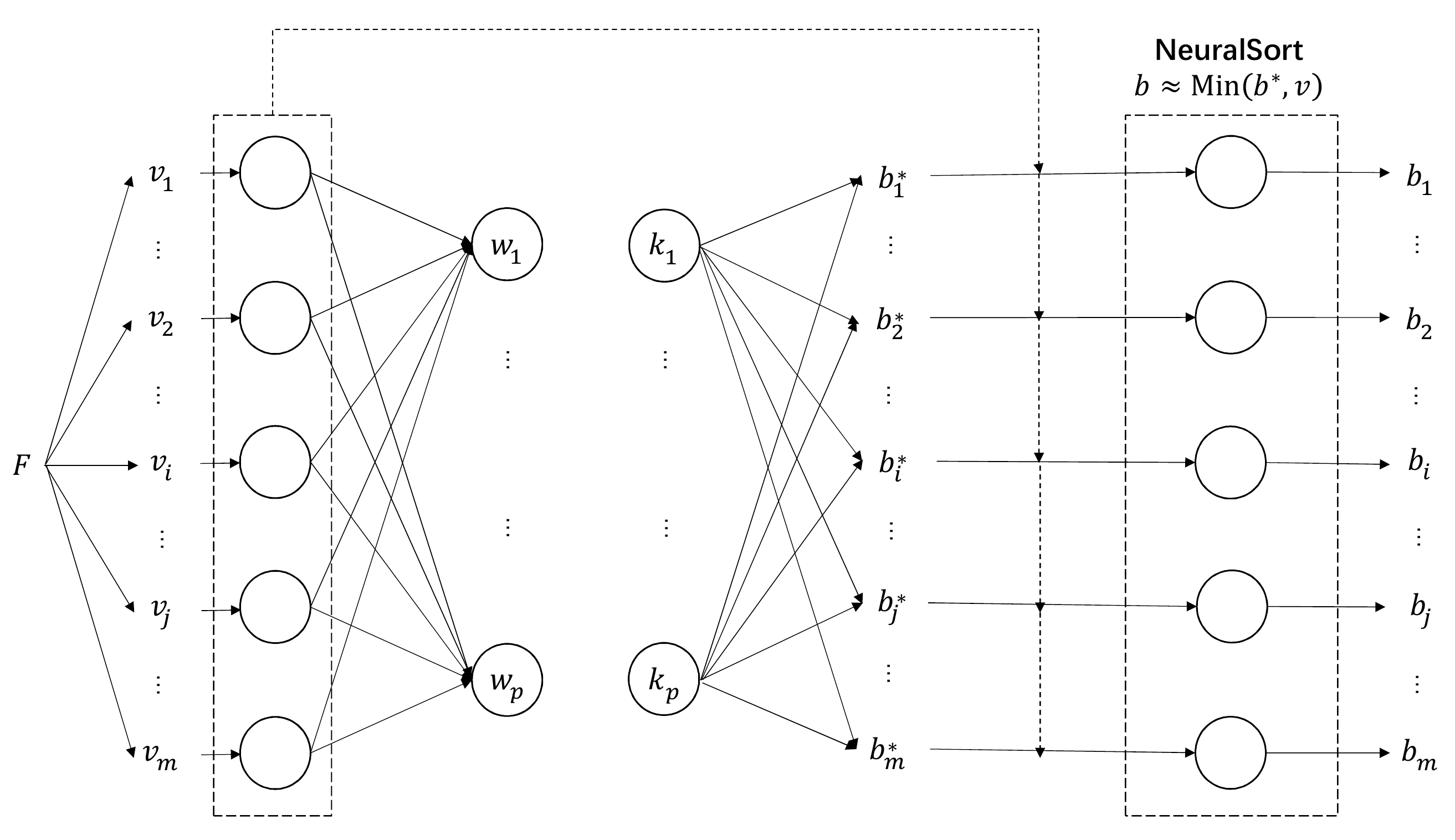}
  \vspace{-0.2cm}
  \caption{The network for strategic bidder (Bid Net), which takes the value of the bidder as input. NeuralSort is a differentiable sorting operator that can output an approximate sorted sequence while preserving the gradient.}
  \label{fig: bid net}
  \Description{Bid Net.}
  \vspace{-0.4cm}
\end{figure}

If we consider static strategies, then for the game induced by the Myerson auction, the Nash equilibrium in the standard setting is:
\begin{equation*}
B_{i}\left(v_{i}\right)= \frac{v_{i} + 1}{4},   \quad g_i(B_i(v_i)) = \frac{1}{2}v_i  , \quad i \in \{1,2\}.
\end{equation*}
However, its drawback is that $B(v) \leq v$ does not always hold. This equilibrium strategy may lead to negative utilities for bidders with $v < \frac{1}{3}$. When bidders do not have access to the dynamic mechanism update method, adopting a bidding strategy that does not satisfy IR (individual rationality) may reduce the bidders' expected utility. Therefore, we require that the bidding strategy obtained by the network satisfies the following requirements:
\begin{itemize}
  \item [1)]
  $B(v)$ is increasing function about $v$,
  \item [2)]
  $v \geq B(v) \geq 0$.
\end{itemize}

We propose a parametric network (Bid Net) to represent the bidder's strategy. The network first inputs the received values $v_i$ from the distribution $F_i$ into a multilayer perceptron. Monotonic outputs $b_i^*=B_i^*(v_i)$ are combined with regularization constraints $B_i(v_i)=\text{Min}[B_i^*(v_i),v_i]$. This ensures that the output of the network meets both of these requirements. Since Bid Net training requires the use of a gradient propagation algorithm and the operators associated with sorting $(Min,Max,Sort)$ cannot propagate the gradient, we use NeuralSort \cite{no21,no22} as an approximation in our network. The structure of the Bid Net is shown in Figure 2.

\subsection{Modeling Repeated Auction as an MARL System}

We assume that bidders and the seller interact in an infinitely repeated auction with identical items and update their strategies after each auction round. At the moment $t$, the seller first announces the current parameter $\theta$ of the mechanism. The strategy of the bidders $\vec{\pi}$ and the seller $\theta$ determines the utility $u^t_i = a^t_iv^t_i-p^t_i$ and the revenue $r^t$ of this auction round. For strategic bidders and the seller, the observations they receive are the joint bidding $\vec{b}$, mechanism parameter $\theta$, and their rewards ($u^t_i$ or $r^t$). When the value distribution is constant, the joint bidding distribution $\vec{b}$ is directly determined by the joint strategy $\vec{\pi}$. Therefore, we use strategies $\vec{\pi}$ instead of bids $\vec{b}$ to represent the action of the bidders in repeated auctions.

Based on the observations and the reward maximization objectives, the agents will update their strategy for the next auction round. We usually assume that the strategic bidder adjusts the previous strategy based on observation: 
\begin{equation*}
    \pi^{t+1}_{i} = \pi^{t}_{i} + \Delta \pi^{t}_{i} (\pi^{t}_{-i}, \theta^t, u^t_i).
\end{equation*}
This forms an MARL system when both the seller and bidders update their strategies through learning. The convergence of the strategy in repeated auctions is consistent with the equilibrium of the induced game, where the bidding strategy and mechanism reach a stable point. Our goal is to provide a learning algorithm for bidders that maximizes their utility as the system converges. When the bidder $i$ is a naive learner \cite{no18}, it will maximize its utility assuming that the other agent strategies remain unchanged:
\begin{equation*}
    \pi^{t+1}_i = \mathop{\text{argmax}}\limits_{\pi_i} U_i(\pi_i, \pi^{t}_{-i}, \theta^t).
\end{equation*}
We can design gradient-based methods with learning rate $\gamma$:
\begin{equation*}
    \Delta \pi^{t}_{i} = \gamma \cdot \nabla_{\pi^{t}_{i}}  U_i(\pi^{t}_{i}, \pi^{t}_{-i}, \theta^t).
\end{equation*}

While the naive learner can only engage in inefficient learning process, the opponent modeling approach is able to improve the reward of agents by predicting the change in the opponent's strategy and selecting the corresponding best response:
\begin{equation*}
    \pi^{t+1}_i = \mathop{\text{argmax}}\limits_{\pi_i} U_i(\pi_i, \pi^{t}_{-i} + \Delta \hat{\pi}_{-i}, \theta^t + \Delta \hat{\theta}).
\end{equation*}
A simple assumption in \cite{no18} is that the other agents are naive learners with lookahead rate $\gamma^\prime$, which means that:
\begin{equation*}
    \Delta \hat{\pi}^{t}_{j} = \gamma^\prime \cdot \nabla_{\pi^{t}_{j}}  U_j(\pi^{t}_{j}, \pi^{t}_{-j}, \theta^t), \ \forall{j \neq i} \ \text{and} \ \Delta \hat{\theta}^t = \gamma^\prime \cdot \nabla_{\theta^{t}}  R(\theta^{t}, \vec{\pi}^{t}).
\end{equation*}

In the following sections, we will specifically discuss the design of the opponent modeling algorithm applicable to repeated auctions and propose our strategy learning method in this MARL system.

\subsection{Opponent Modeling Based Automatic Bidding Method}

In this section, we introduce the automatic bidding method for the MARL system in Section 3.2. Since the seller and other bidders also update their strategies, agents must consider both the current utility and the impact on subsequent states when evaluating the strategy $\pi_i$. We adopt representation similar to RL (reinforcement learning) and use $Q_i$ to denote the utility expectation of the strategy for the bidder $i$, then we have the following:
\begin{equation*}
    Q_i(\pi^t_i) = U_i(\pi^t_i,\pi^t_{-i},\theta^t) + \lambda \mathop{\text{max}}\limits_{\pi_i} Q_i(\pi_i,\pi^{t+1}_{-i},\theta^{t+1}),
\end{equation*}
where $\lambda$ is the discount factor.

\begin{figure}[]
  \centering
  \includegraphics[width=0.8\linewidth]{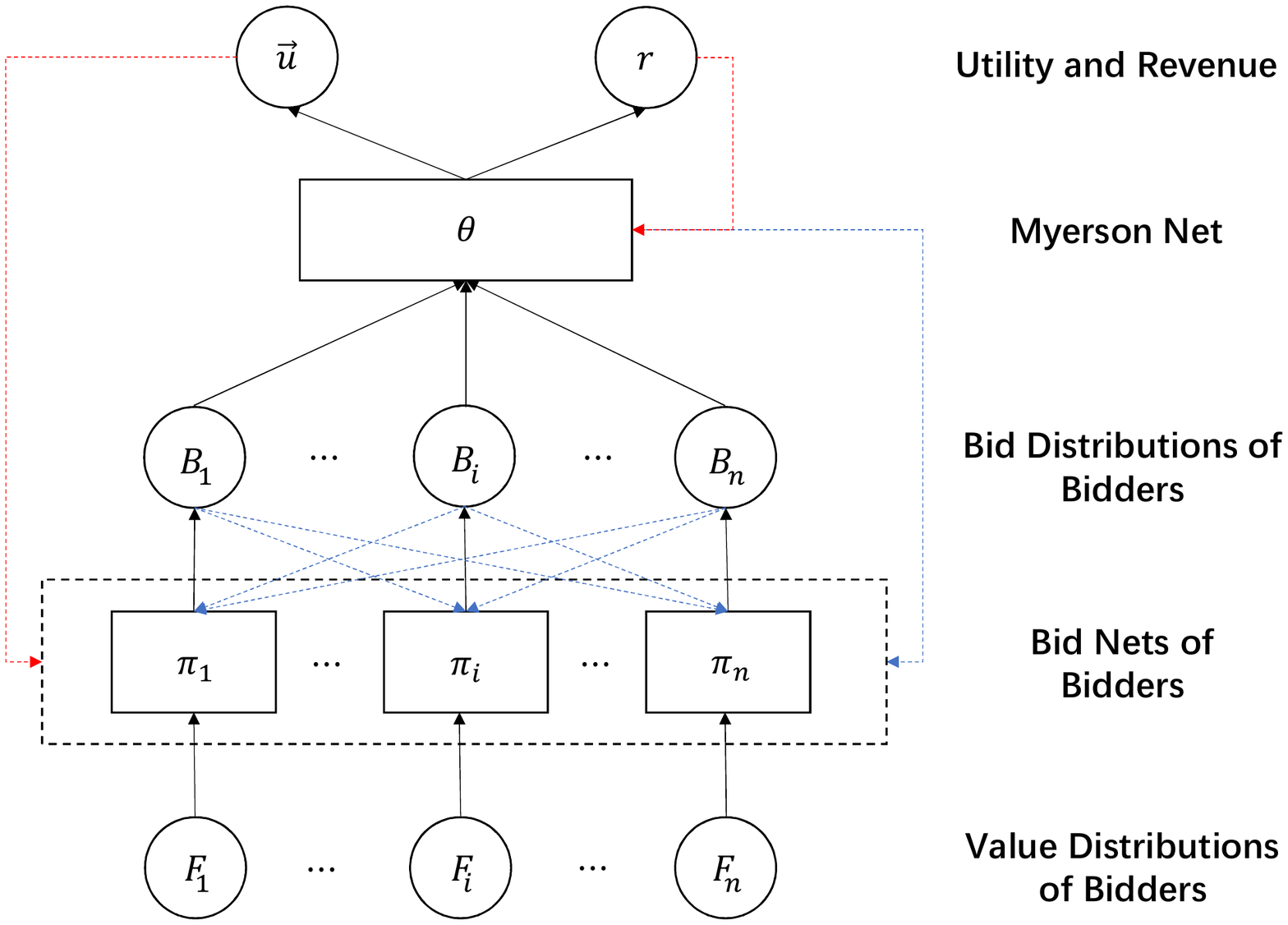}
  \caption{Gradient propagation direction of the repeated auction induced MARL system. The red line represents a direct gradient, which comes from the revenue and utility. The blue line represents the indirect gradient, which comes from the impact of the bidder's strategy on other players.}
  \label{fig: bid net}
  \Description{Bid Net.}
  \vspace{-0.4cm}
\end{figure}

For the agent $i$ who needs to choose a strategy at moment $t$, the information available is historical observations before moment $t$. Therefore, it is difficult to obtain unbiased estimates of $Q_i$. A direct approximation is $\hat{Q}_i = U_i(\pi^t_i,\pi^t_{-i},\theta^t)$, which means that the bidder assumes that the subsequent strategy of other agents ($\pi^{t+1}_{-i}, \theta^{t+1}$) is not relevant to its strategy $\pi^t_i$. Then its choice of action at the next moment will be based on the prediction of the opponent's action:
\begin{equation*}
\begin{aligned}
   & \pi^{t}_i = \mathop{\text{argmax}}\limits_{\pi_i} U_i(\pi_i,\hat{\pi}^{t}_{-i},\hat{\theta}^{t}), \\
    \hat{\pi}^{t}_{-i}=  \hat{\pi}^{t}_{-i}(\pi^{t-1}_i & ,\pi^{t-1}_{-i},\theta^{t-1}), \quad \hat{\theta}^{t} = \hat{\theta}^{t}(\pi^{t-1}_i,\pi^{t-1}_{-i},\theta^{t-1}).
\end{aligned}
\end{equation*}

For the naive learner, it will assume that the strategies of other agents remain unchanged:
\begin{equation*}
\hat{\pi}^{t}_{-i} = \pi^{t-1}_{-i} , \quad \hat{\theta}^{t} = \theta^{t-1}.
\end{equation*}
Then its gradient update direction for the strategy will be:
\begin{equation*}
\nabla_{\pi^{t-1}_i}U_i(\pi^{t-1}_i,\pi^{t-1}_{-i},\theta^{t-1}).
\end{equation*}

More accurate estimations for $U_i(\pi^t_i,\pi^t_{-i},\theta^t)$ usually require first-order approximations:
\begin{equation*}
\begin{aligned}
U_{i}(\pi_{i}, &  \pi_{-i}+\Delta \hat{\pi}_{-i}, \theta+\Delta \hat{\theta}) \\
& \approx U_{i}(\pi_{i}, \pi_{-i}+\Delta \hat{\pi}_{-i}, \theta)+(\Delta \hat{\theta})^{\top} \nabla_\theta U_{i}(\pi_{i}, \pi_{-i}+\Delta \hat{\pi}_{-i},\theta) \\
& \approx U_{i}(\pi_i,\pi_{-i},\theta) + (\Delta \hat{\pi}_{-i})^{\top} \nabla \pi_{-i} U_{i}(\pi_i,\pi_{-i},\theta) \\
& + (\Delta \hat{\theta})^{\top} \nabla_\theta [U_{i}(\pi_i,\pi_{-i},\theta) + (\Delta \hat{\pi}_{-i})^{\top} \nabla \pi_{-i} U_{i}(\pi_i,\pi_{-i},\theta)].
\end{aligned}
\end{equation*}

We can obtain different predictions for $(\Delta \hat{\pi}_{-i},\Delta \hat{\theta})$ with opponent modeling methods. The LOLA \cite{no18} algorithm assumes that the opponent is a naive learner, which means $\Delta \hat{\theta} = \eta \cdot \nabla_\theta R(\vec{\pi},\theta)$. Assuming that the opponent is a LOLA learner leads to high-order LOLA (HOLA). Similar opponent modeling algorithms include SOS (stable opponent shaping) \cite{no19}, COLA (consistent learning with opponent learning awareness) \cite{no20}, and others, which effectively learn equilibrium with higher rewards in different game environments.

COLA \cite{no20} points out that the effectiveness of opponent modeling algorithms depends on the consistency of the lookahead rate of all agents. This means that agents need to choose strategy-update approaches with consistency to ensure the convergence of the system. However, we have discussed earlier the asymmetry of the repeated auction environment: bidders have private information (value function) that allows them to adopt specific strategies to improve utility. Training strategies directly using opponent modeling algorithms does not maximize the utility of bidders.

Figure 3 shows the direction of propagation of the network parameter gradient for bidders and the seller in the MARL environment of repeated auctions. We illustrate the necessity of computing the indirect gradient represented by the blue dashed line in the figure in repeated auctions by proving the following property.
\begin{theorem}
If all bidders use only the gradient which comes directly from the utility to update the strategy under the Myerson Net of the seller, their gradient-based strategy updating will converge to the truthful bidding strategy $B_i(v_i)=v_i$. The stable state of the system is where all bidders bidding truthfully.
\end{theorem}

The proof of Theorem 3.1 is given in the supplementary material. This Theorem indicates that the strategy network of bidder must be trained with the indirect gradient, which requires predicting the impact of bidding strategies on mechanism changes. In order to obtain a more accurate prediction of the $Q_i$ function for bidders, we estimate $\mathop{\text{max}}\limits_{\pi_i} Q_i(\pi_i,\pi^{t+1}_{-i},\theta^{t+1})$ by modeling $\Delta \hat{\pi}^{t+1}_{-i},\Delta \hat{\theta}^{t+1}$ as functions of $\pi^t_i$:
\begin{equation*}
    \Delta \hat{\pi}^{t+1}_{-i} = \Delta \hat{\pi}^{t+1}_{-i}(\pi^t_i), \quad  \Delta \hat{\theta}^{t+1} = \Delta \hat{\theta}^{t+1}(\pi^t_i).
\end{equation*}
Then we have:
\begin{equation*}
\begin{aligned}
    \hat{Q}_i(\pi^t_i,\pi^{t+1}_i,\cdots) = & U_i(\pi^t_i, \hat{\pi}^t_{-i}, \hat{\theta}^t) + \lambda U_i(\pi^{t+1}_i, \hat{\pi}^t_{-i} + \Delta \hat{\pi}_{-i}(\pi^t_i), \hat{\theta}^t \\
    + &  \Delta \hat{\theta}(\pi^t_i)) + \lambda^2 \cdots.
\end{aligned}
\end{equation*}

The convergence of the system in the repeated auction is equivalent to the convergence of the strategies of each agent. We can simplify the above equation by assuming that $\pi^t_i = \pi^{t+1}_i = \cdots$ and then it will become:
\begin{equation*}
    \hat{Q}_i(\pi^t_i) = U_i(\pi^t_i, \hat{\pi}^t_{-i}, \hat{\theta}^t) + \sum^\infty_{k=1} \lambda^k U_i(\pi^t_i, \hat{\pi}^t_{-i} + \Delta^k \hat{\pi}_{-i}(\pi^t_i), \hat{\theta}^t +  \Delta^k \hat{\theta}(\pi^t_i)),
\end{equation*}
where $\Delta^k \hat{\pi}_{-i}(\pi^t_i)$ means the prediction of $\pi_{-i}$ after $k$ updates.

We use $\hat{\pi}_{-i}(\pi^t_i), \hat{\theta}(\pi^t_i)$ to represent the strategy of other agents when the system converges, which means that:
\begin{equation*}
    \hat{\pi}_{-i}(\pi^t_i) = \hat{\pi}^t_{-i} + \Delta^\infty \hat{\pi}_{-i}(\pi^t_i), \quad \hat{\theta}(\pi^t_i) = \hat{\theta}^t +  \Delta^\infty \hat{\theta}(\pi^t_i).
\end{equation*}

We can always require that the system converge in finite time by periodic reducing the step size of the strategy updating. Assuming that the system converges after $T$ updates, we have:
\begin{equation*}
    \hat{Q}(\pi^t_i) = \sum^{T-1}_{k=0} U^{k+t}_i + (\lambda^T + \lambda^{T+1} + \cdots) U_i(\pi^t_i, \hat{\pi}_{-i}(\pi^t_i), \hat{\theta}(\pi^t_i)).
\end{equation*}

When $\lambda \rightarrow 1$, the second item will be sufficiently larger than the first. Then we have:
\begin{equation*}
    \hat{Q}(\pi^t_i) \approx (\lambda^T + \lambda^{T+1} + \cdots) U_i(\pi^t_i, \hat{\pi}_{-i}(\pi^t_i), \hat{\theta}(\pi^t_i)).
\end{equation*}
Thus, the strategy selection function of the agent is as follows:
\begin{equation*}
    \pi^t_i =  \mathop{\text{argmax}}\limits_{\pi_i} U_i(\pi_i, \hat{\pi}_{-i}(\pi^t_i), \hat{\theta}(\pi^t_i)).
\end{equation*}

When a single bidder's strategy is fixed in a repeated auction, the strategy updates of other bidders and the seller are synchronized and will affect each other. The convergence result is determined by the algorithm used by these agents. To avoid discussing the possibility of different stable points in the system, we assume that the strategy updates of other agents are independent and the bidders' strategies change slowly $\Delta \hat{\pi}_{-i}(\pi^t_i) \approx 0$. We found this assumption to be valid in our experiments, for predicting the strategies of other agents under the condition that their private information is unknown will lead to a large bias.

We refer to this process (calculating $\hat{\pi}_{-i}(\pi^t_i), \hat{\theta}(\pi^t_i)$ using $\pi^t_i$) as the inner loop part of the algorithm. When the seller adopts Myerson Net as the mechanism, its strategy updating process according to historical bids is similar to a naive learner. We can simulate this process in the inner loop through the Myerson Net by constraining the bidder strategies to $\pi^t_i,\pi^{t-1}_{-i}$. The complete procedure for the inner loop is given in Algorithm 1.

\begin{algorithm}[t]
\caption{Inner loop process of pseudo-gradient algorithm}\label{algo1}
\begin{algorithmic}[1]
        \REQUIRE Initial Parameters $\pi^{t-1}_i , \Delta \pi_i, \pi^{t-1}_{-i}, \theta^{t-1}$; termination step $T$; lookahead rate $\eta^\prime$.
        \STATE Generate Myerson Net with parameter $\theta^*$ which statisfies $M(\theta^*) \cong M(\theta^{t-1})$, and set $\pi^*_{-i} = \pi^{t-1}_{-i}$ 
        \FOR{$k := 1, \cdots , T$}
            \STATE Generate sample $S_k = (b_1, \cdots , b_n)$ with $(\pi^{t-1}_i + \Delta \pi_i, \pi^*_{-i})$
            \STATE Compute gradient $\nabla_\theta R(\theta^*,S_k)$
            \STATE $\theta^* = \theta^* + \eta^\prime \nabla_\theta R(\theta^*,S_k), \ \pi^*_{-i} = \pi^*_{-i}$
        \ENDFOR
        \ENSURE $\hat{\theta}^t = \theta^*, \hat{\pi}^t_{-i} = \pi^*_{-i}$
\end{algorithmic}
\end{algorithm}

Based on the strategy predictions of the other agents obtained from the inner loop, we can derive the bidding strategy that maximizes the expected utility. To avoid the instability caused by rapid strategy changes, we restrict the step size by $\lvert \Delta \pi_i \rvert \leq d$. Then our goal is to solve the optimization problem:
\begin{equation*}
 \mathop{\text{max}}\limits_{\lvert \Delta \pi_i \rvert \leq d} \  U_i(\pi_i + \Delta \pi_i, \hat{\pi}_{-i}(\pi_i + \Delta \pi_i), \hat{\theta}(\pi_i + \Delta \pi_i))
\end{equation*}

Since the parameters in the optimization objective contain the output of the inner loop, it is difficult to calculate the strategy gradient directly. We give the approximate calculation based on the PG (pseudo-gradient) by defining the pseudo-gradient obtained from $\Delta \pi_i$ as:
\begin{equation*}
\begin{aligned}
grad(\pi_i, \Delta \pi_i)  
 = & \frac{U_i(\pi_i + \Delta \pi_i,\pi_{-i} + \Delta \hat{\pi}_{-i}(\Delta \pi_i),\theta + \Delta \hat{\theta}(\Delta \pi_i))}{\lvert \Delta \pi_i \rvert} \\
 - & \frac{U_i(\pi_i,\pi_{-i},\theta)}{\lvert \Delta \pi_i \rvert}. 
\end{aligned}
\end{equation*}

From the definition we can see that, given the strategy update $\Delta \pi_i$, it is possible to calculate the pseudo-gradients with the inner loop. For the algorithm to choose an update step that is close to the direction of the true gradient, we generate a set of different directions of $\Delta \pi_i$ and select the positive gradient with the largest absolute value from all pseudo-gradients as the update direction of the Bid Net. The complete procedure for the PG algorithm is given in Algorithm 2. We want the algorithm to converge to the equilibrium of the repeated auction-induced game. For a single-item Myerson auction that induces a game among bidders, we give proof of the convergence in the supplementary material.

\begin{theorem}
Assuming that other bidders use static strategies $\pi_{-i}$, $\hat{\theta}(\Delta \pi_i)$ obtained from the inner loop belonging to the optimal mechanisms and $K^\prime \rightarrow \infty$ in algorithm 2, the strategic bidder using the PG algorithm will converge to the strategy with optimal utility. When all bidders adopt the PG algorithm, the equilibrium of the Myerson auction-induced game is the only stable point of the system.
\end{theorem}

In our experiments, we find that the result of the algorithm satisfies this theorem even if the inner loop does not converge and $K^\prime$ is finite. Our algorithm PG converges to the equilibrium of all the games induced by the auction, even in multiple-bidders settings.

\begin{algorithm}[t]
\caption{Pseudo-gradient algorithm}\label{algo2}
\begin{algorithmic}[1]
	\REQUIRE Initial Parameters $\pi^0_i , \pi^0_{-i}, \theta^0$, termination step $T^\prime$, learning rate $\eta$, number of inner loop $K^\prime$ and constant $s$ and $l$.
        \FOR{$t := 1, \cdots , T^\prime$}
            \STATE Strategic bidder process: read the output of $t-1$ from environment and set $\hat{\pi}_{-i} = \pi^{t-1}_{-i}, \hat{\theta} = \theta^{t-1}$
            \FOR{$j := 1, \cdots , K^\prime$}
                \STATE Randomly generate $\Delta \pi^j_i$, which satisfies $	\langle \Delta \pi^j_i, \Delta \pi^{j_0}_i \rangle < l  \ (\forall j_0 \in \{ j-1,\cdots,j-s \})$
                \STATE Using the inner loop process to obtain $\hat{\theta}(\Delta \pi^j_i)$
            \ENDFOR
            \STATE $grad(\pi^{t-1}_i, \Delta \pi^j_i) = \frac{U_i(\pi^{t-1}_i + \Delta \pi^j_i,\hat{\pi}_{-i},\hat{\theta}) - U_i(\pi^{t-1}_i,\pi_{-i},\theta)}{\lvert \Delta \pi^j_i \rvert}   $
            \STATE  $grad^*(\pi^{t-1}_i) = \mathop{\text{argmax}}\limits_{ \Delta \pi^j_i} [grad(\pi^{t-1}_i, \Delta \pi^j_i) \mid grad(\cdot , \cdot)>0]$
            \STATE $\pi^t_i = \pi^{t-1}_i + \eta \cdot grad^*(\pi^{t-1}_i)$
            \STATE Other bidders and the seller update their strategies according to their algorithms
        \ENDFOR
	\ENSURE $(\pi^{T^\prime}_i , \pi^{T^\prime}_{-i}, \theta^{T^\prime})$
\end{algorithmic}
\end{algorithm}


\begin{figure*}[]
  \centering
  \subfigure[Bid Net with RL training]{\includegraphics[width=0.25\linewidth]{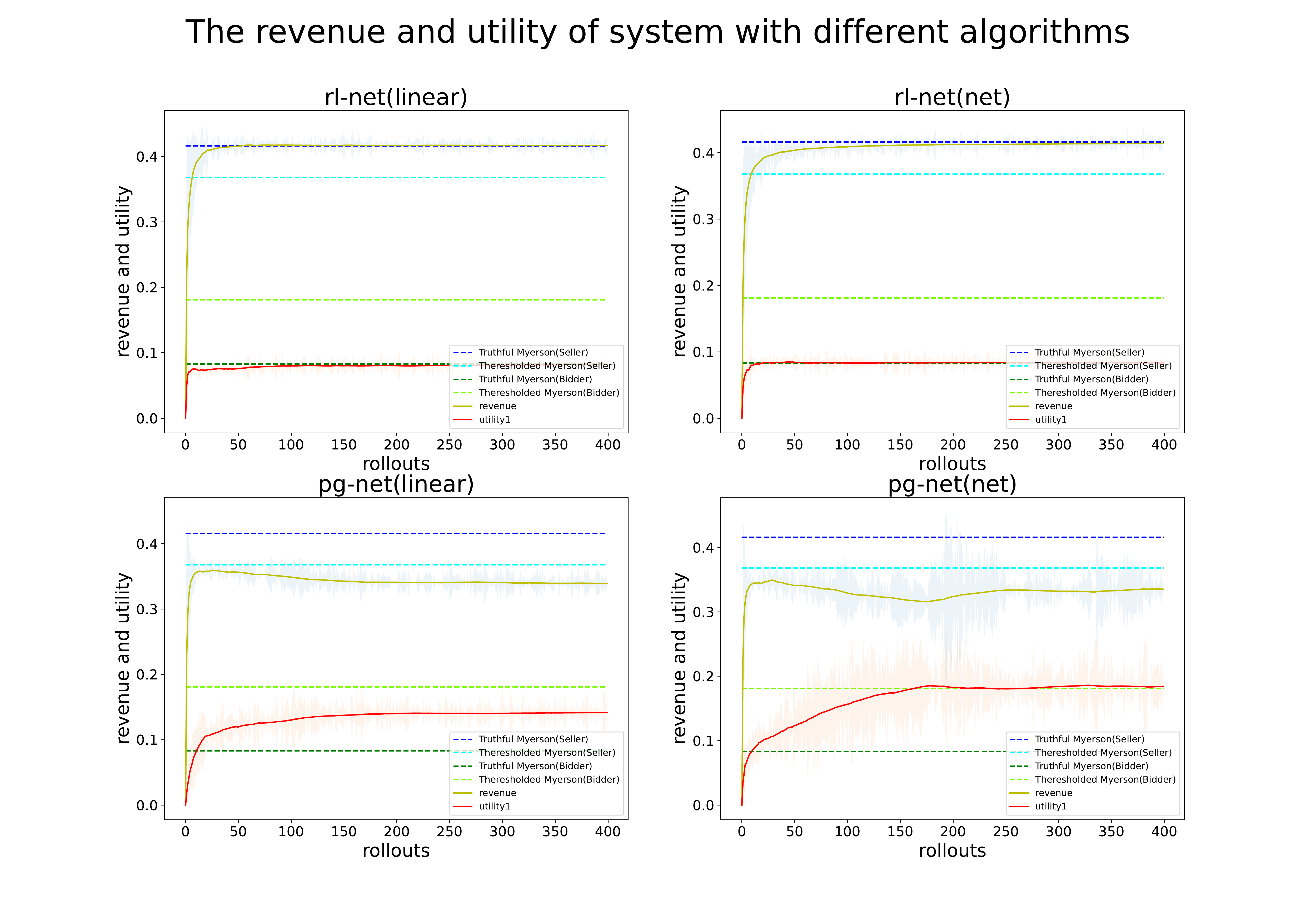}}
  \subfigure[Linear strategy with PG training]{\includegraphics[width=0.25\linewidth]{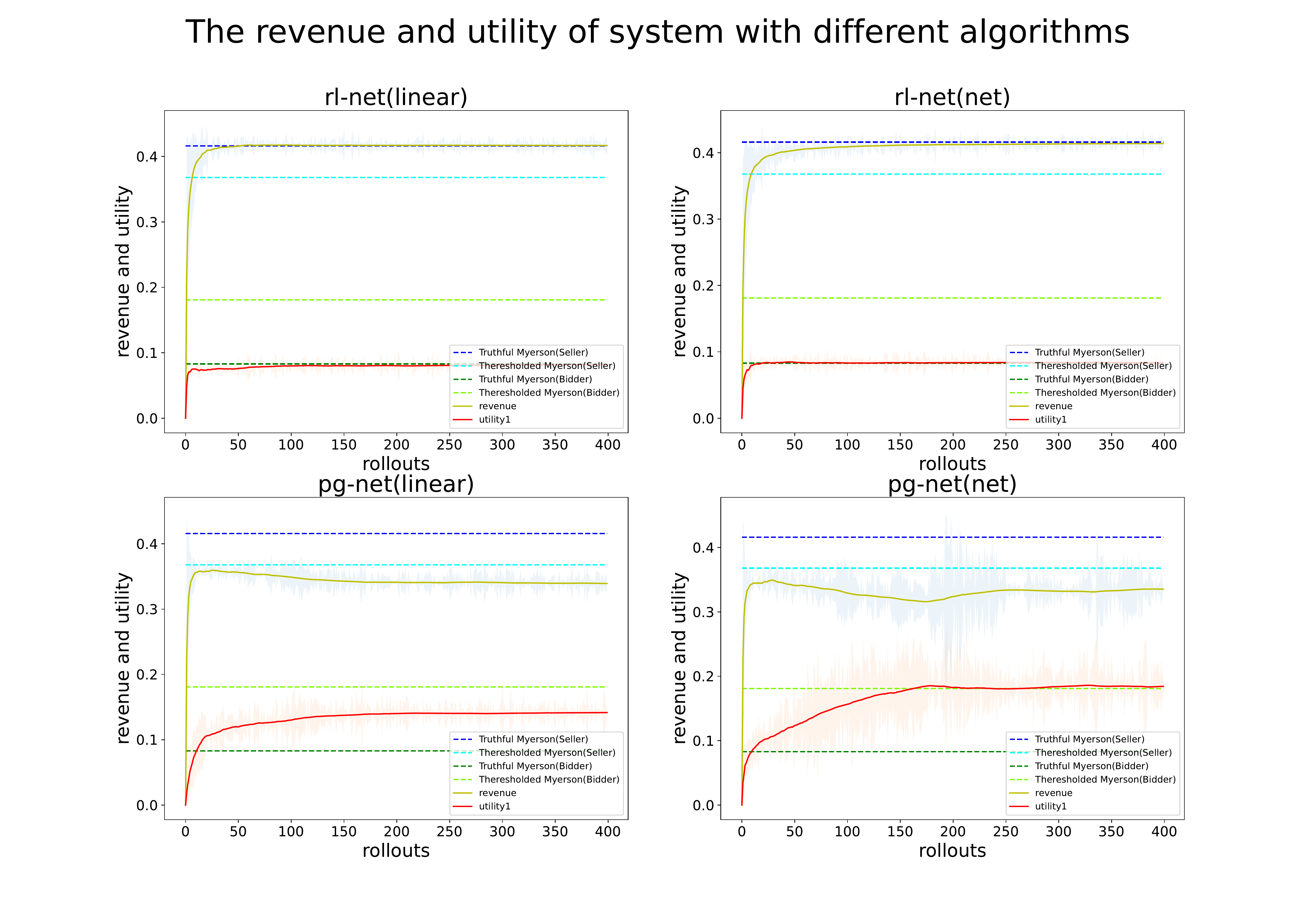}}
  \subfigure[Bid Net with PG training]{\includegraphics[width=0.25\linewidth]{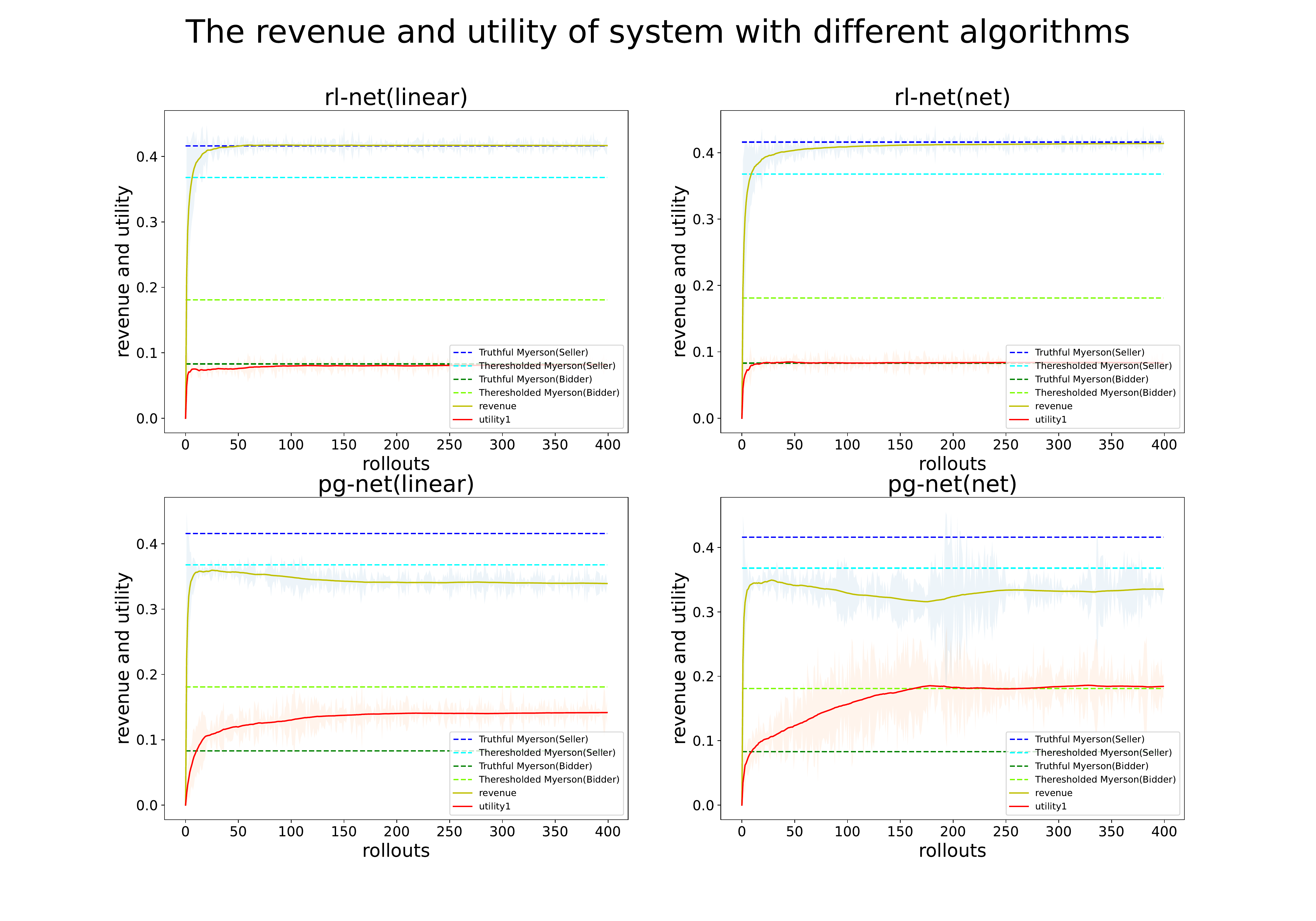}}
  \vspace{-0.3cm}
\caption{The utility of a strategic bidder in scenarios where another bidder consistently employs the truthful bidding strategy, while the seller's strategy is derived from the Myerson Net. The solid red line represents the utility of the strategic bidder, while the solid yellow line represents the revenue of the seller. The dashed line labeled "Truthful Myerson" represents the theoretical utility and revenue when the strategic bidder adheres to the truthful bidding strategy. The dashed line labeled "Theresholded Myerson" illustrates the theoretical utility and revenue when the strategic bidder employs the optimal bidding strategy.}
\vspace{-0.3cm}
\end{figure*}

\begin{figure*}[]
  \centering
  \subfigure[rl]{\includegraphics[width=0.2\linewidth]{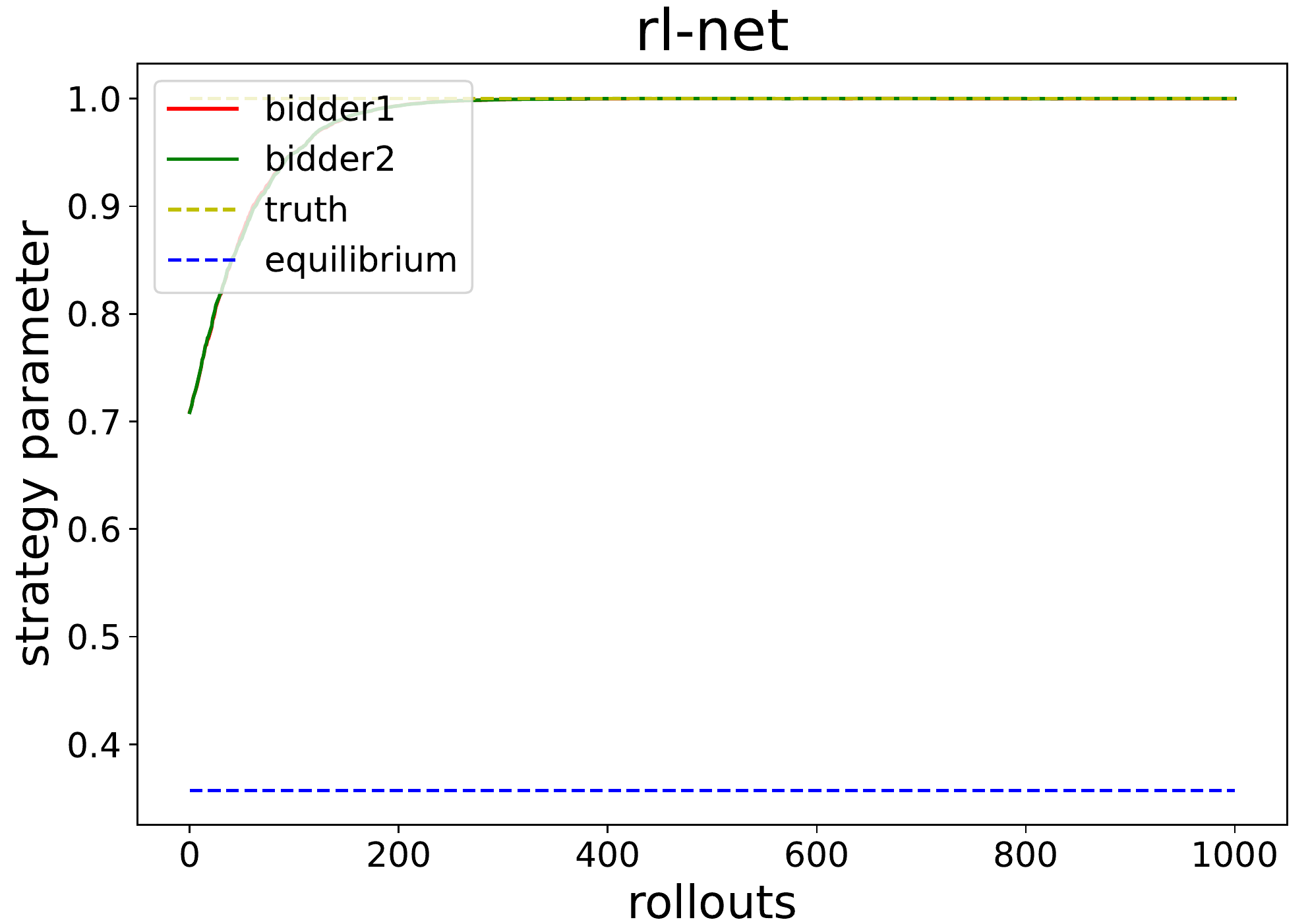}}
  \subfigure[lola]{\includegraphics[width=0.2\linewidth]{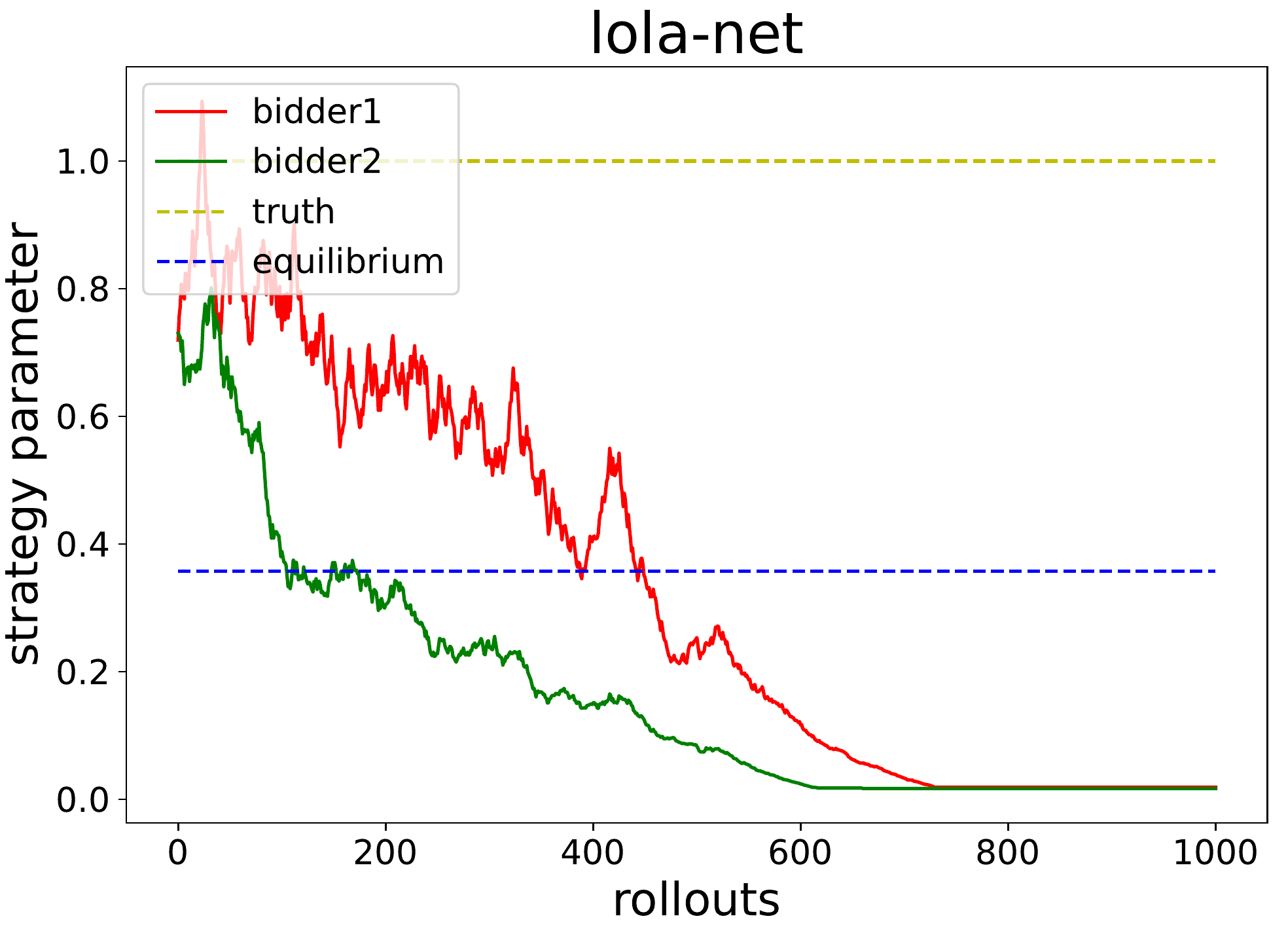}}
  \subfigure[la]{\includegraphics[width=0.2\linewidth]{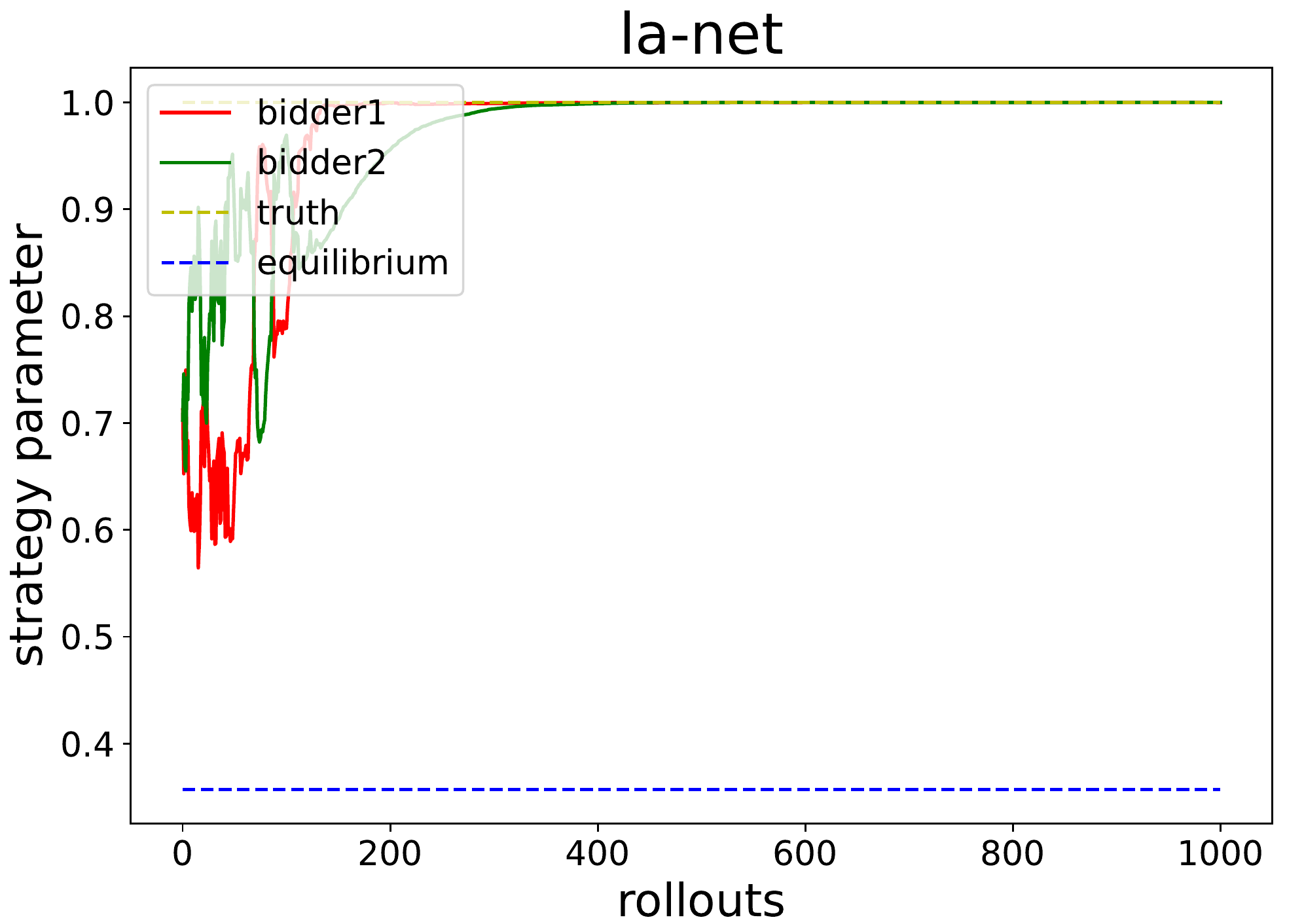}}
  \subfigure[co]{\includegraphics[width=0.2\linewidth]{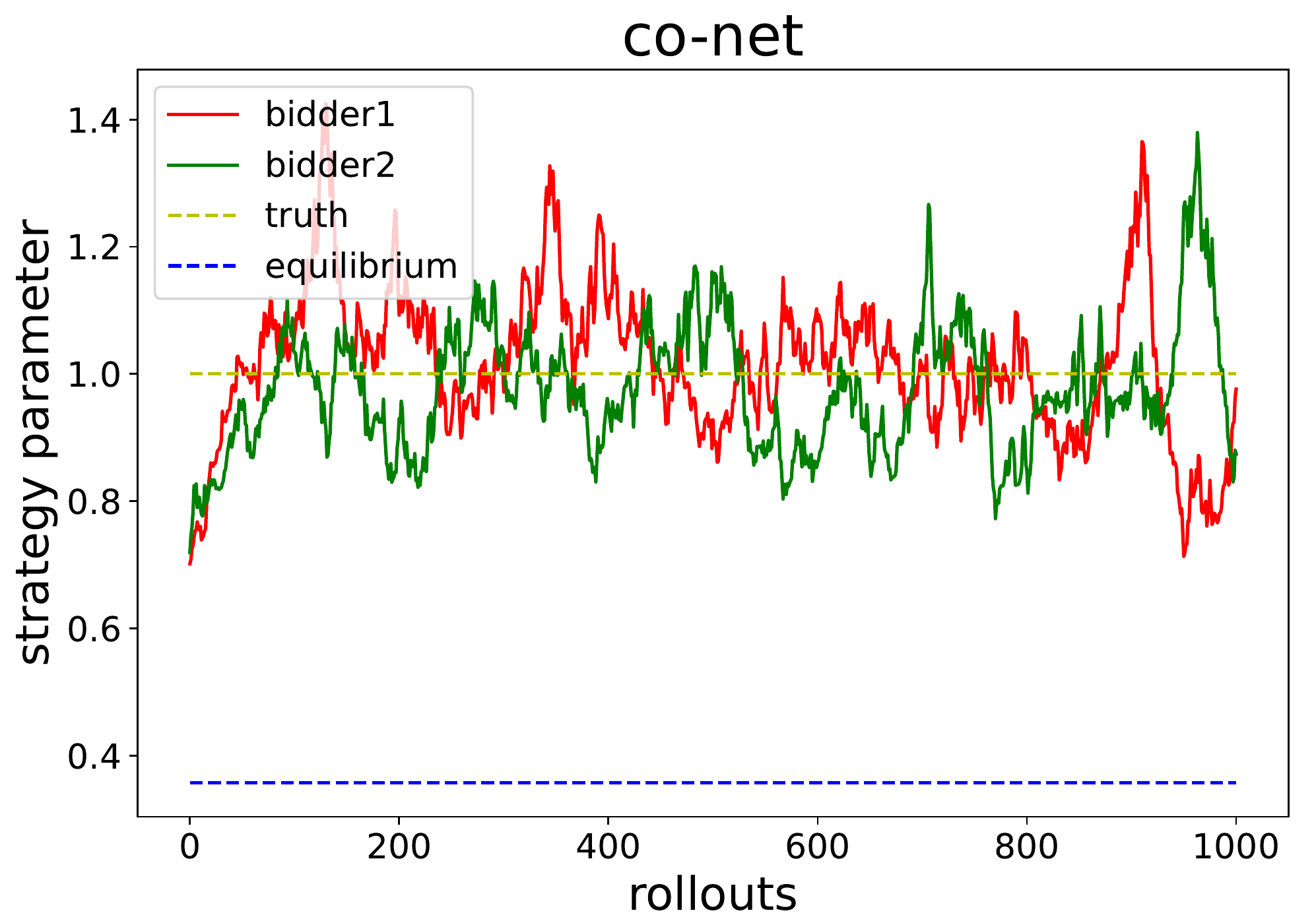}}
  \subfigure[cgd]{\includegraphics[width=0.2\linewidth]{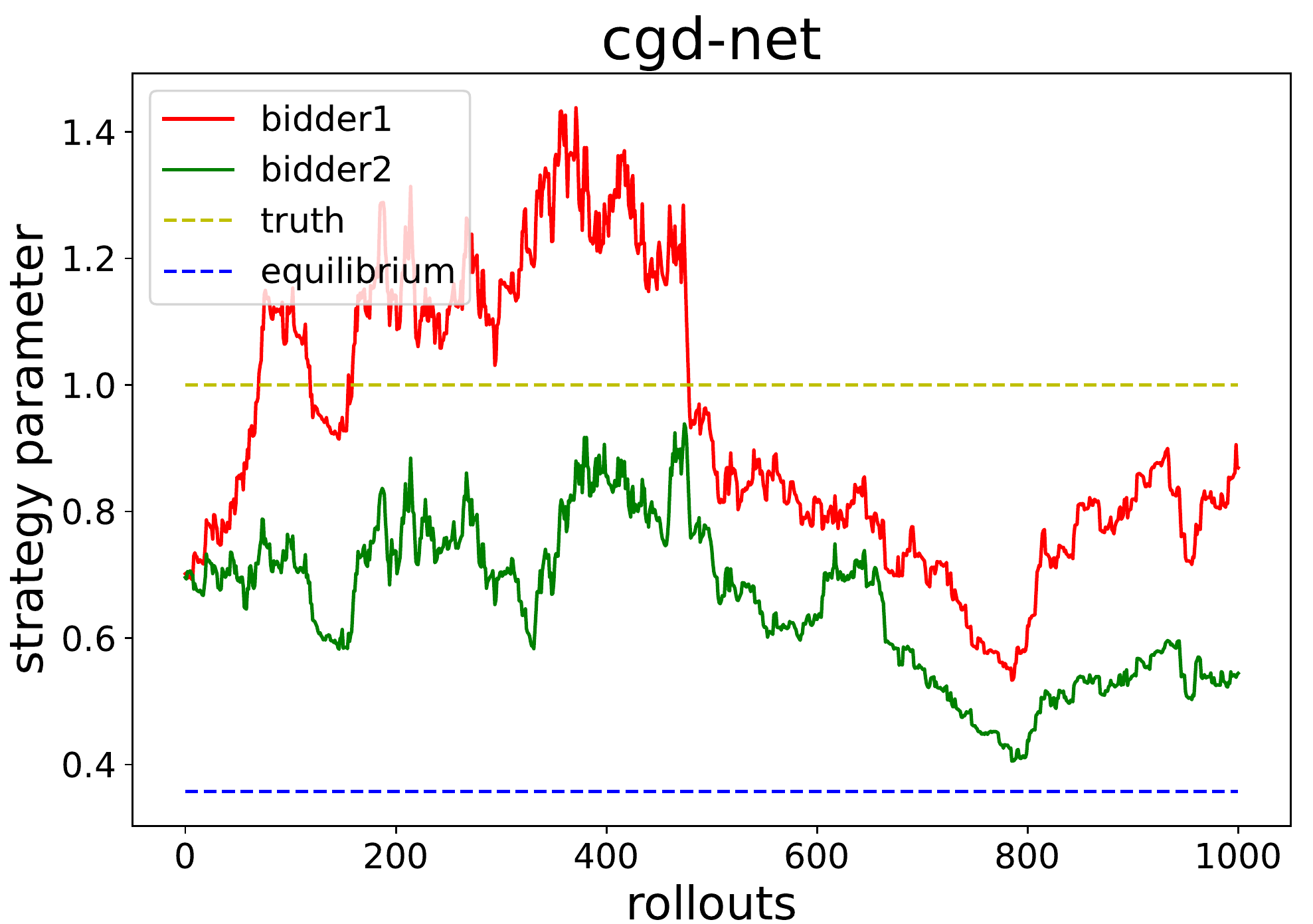}}
  \subfigure[lss]{\includegraphics[width=0.2\linewidth]{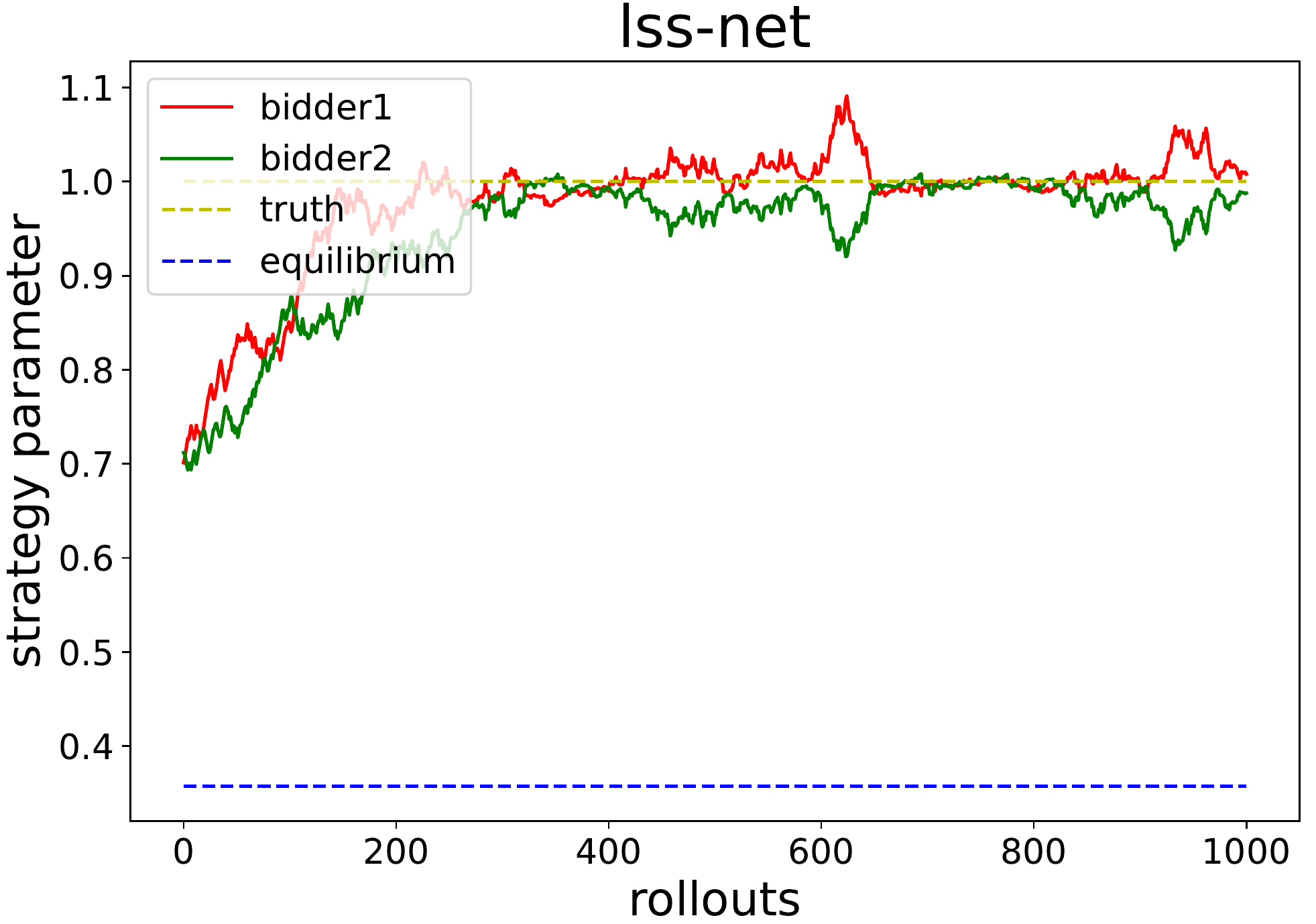}}
  \subfigure[sos]{\includegraphics[width=0.2\linewidth]{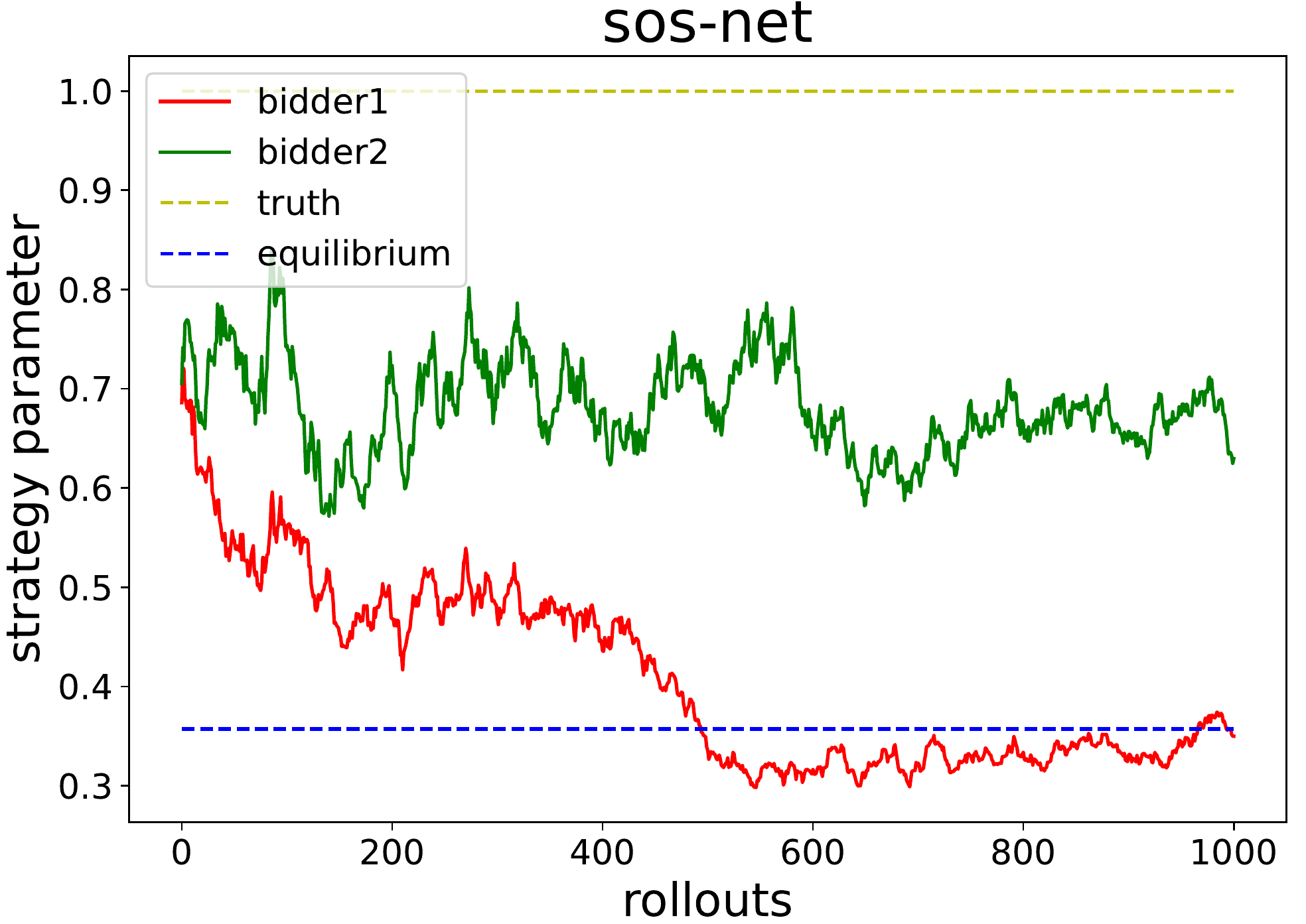}}
  \subfigure[pg]{\includegraphics[width=0.2\linewidth]{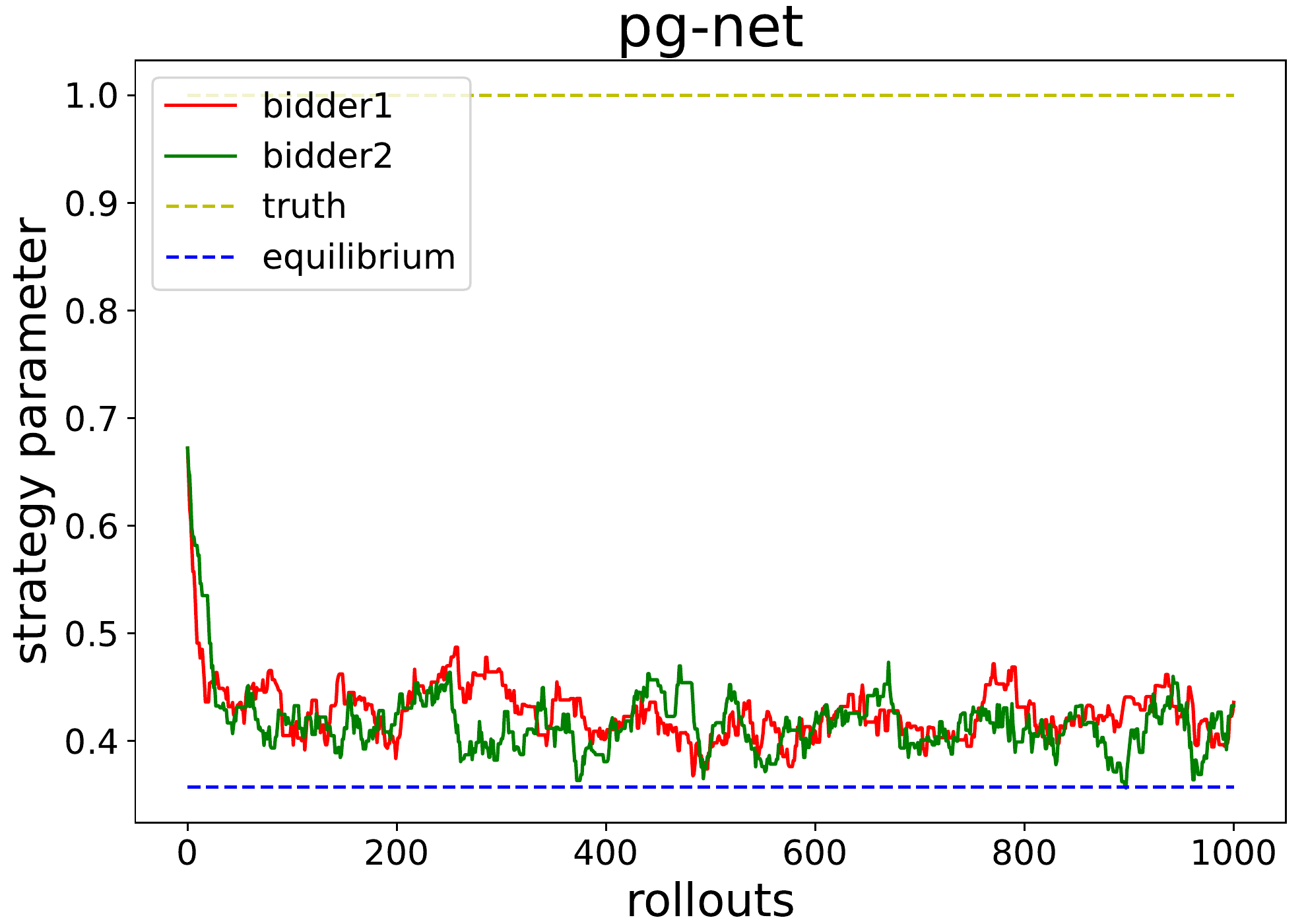}}
  \vspace{-0.3cm}
\caption{The strategy parameters $\alpha_i$ of strategic bidders during the learning process. Both bidders are designated strategic bidders and employ the same algorithm. The seller strategy is derived from Myerson Net. The line labeled "truth" represents the truthful strategy $\alpha_i = 1$, and the line labeled "equilibrium" represents the equilibrium strategy of the induced game $\alpha_i = \frac{5}{14}$. }
\vspace{-0.4cm}
\end{figure*}

\section{Experiments and discussion}

Our experiment consists of three parts. 1) Experiment in the standard setting for Bid Net employed by individual bidders. We compare the learned Bid Net strategies with linear shading strategies \cite{no11}. In this configuration, another bidder consistently employs the truthful bidding strategy, while the seller's strategy is derived from the Myerson Net. 2) Experiment with the PG algorithm. As part of an ablation experiment, we utilize linear shading networks to represent bidder strategies. We compare the PG algorithm with other opponent modeling algorithms and evaluate performance using metrics such as the average utility of the bidders and the deviation of strategy from equilibrium in the induced game. 3) Experiments in diverse environments with varied settings and opponent models. We expand our experiment to include various environments that feature distinct settings and opponent models. Our aim is to demonstrate the effectiveness of the combination of PG algorithm and Bid Net in optimizing bidders' utility in online auctions.

\subsection{Experiment for Bid Net}

We compare linear shading strategies and Bid Net, which is trained using PG and RL algorithms. Our evaluation takes place in the standard setting, with another bidder consistently employing the truthful bidding strategy. The results are given in Figure 4.

From the figure, we can see that the network trained using RL converges to the truthful bidding strategy. The linear shading strategy trained through PG improves the utility of the strategic bidder, but does not approach the optimal strategy. The strategy produced by Bid Net trained with the PG algorithm closely approximates the optimal strategy. This experiment underscores the efficiency of Bid Net in accurately representing bidding strategies, while linear shading strategies do not contain the optimal strategy. The gap between the seller's revenue and the theoretical value may arise from the fact that we set a very small reserve price.

\begin{table*}[]
\resizebox{0.6\linewidth}{!}{
\begin{tabular}{|c|c|c|c|c|c|}
\hline
Setting & RL & LOLA & SOS & PG (linear shading) & PG (Bid Net) \\ \hline
\begin{tabular}[c]{@{}c@{}} \ding{172} $B_2(v_2)= v_2$ \\  $v_1 \sim U[0,1]$, $v_2 \sim U[0,1]$ \end{tabular} & 0.09    & 0.15   & 0.12            & 0.18   & 0.20  \\ \hline
\begin{tabular}[c]{@{}c@{}}\ding{173} $B_2(v_2)=0.25v_2+0.25$\\ $v_1 \sim U[0,1]$, $v_2 \sim U[0,1]$\end{tabular}  & 0.09   & 0.07                     & 0.10   & 0.12    & 0.16    \\ \hline
\begin{tabular}[c]{@{}c@{}}\ding{174} $B_2(v_2)= v_2$\\ $v_1 \sim U[0,2]$, $v_2 \sim U[0,1]$\end{tabular} & 0.22 & 0.37 & 0.33  & 0.33  & 0.46\\ \hline
\begin{tabular}[c]{@{}c@{}}\ding{175} $B_2(v_2)= v_2$\\ $v_1 \sim (\frac{n}{N}+1) \cdot U[0,1]$, $v_2 \sim U[0,1]$\end{tabular}  & 0.16 & 0.26  & 0.09 & 0.27   & 0.33 \\ \hline
\begin{tabular}[c]{@{}c@{}}\ding{176} $B_2(v_2)= v_2$, $B_3(v_3)= v_3$\\ $v_1 \sim U[0,1]$, $v_2 \sim U[0,1]$, $v_3 \sim U[0,1]$\end{tabular}  & 0.06   & 0.09 & 0.08 & 0.04  & 0.11  \\ \hline
\begin{tabular}[c]{@{}c@{}}\ding{177} $B_2(v_2) \sim$ LOLA updating\\ $v_1 \sim U[0,1]$, $v_2 \sim U[0,1]$\end{tabular} & 0.12   & 0.16  & 0.16 & 0.20  & 0.31  \\ \hline
\begin{tabular}[c]{@{}c@{}}\ding{178} $B_2(v_2) \sim$ SOS updating\\ $v_1 \sim U[0,1]$, $v_2 \sim U[0,1]$\end{tabular}  & 0.12  & 0.13   & 0.15 & 0.18  & 0.32  \\ \hline
\end{tabular}}
\caption{The utility of strategic bidder across different opponent strategies and environments. Row labels signify the environment and strategy of the other bidders. The column labels indicate the algorithms employed by the strategic bidder. $n$ and $N$ denote the number of current rollouts and the total number of rollouts. }
\vspace{-0.5cm}
\end{table*}

\begin{figure}[]
  \centering
  \includegraphics[width=0.7\linewidth]{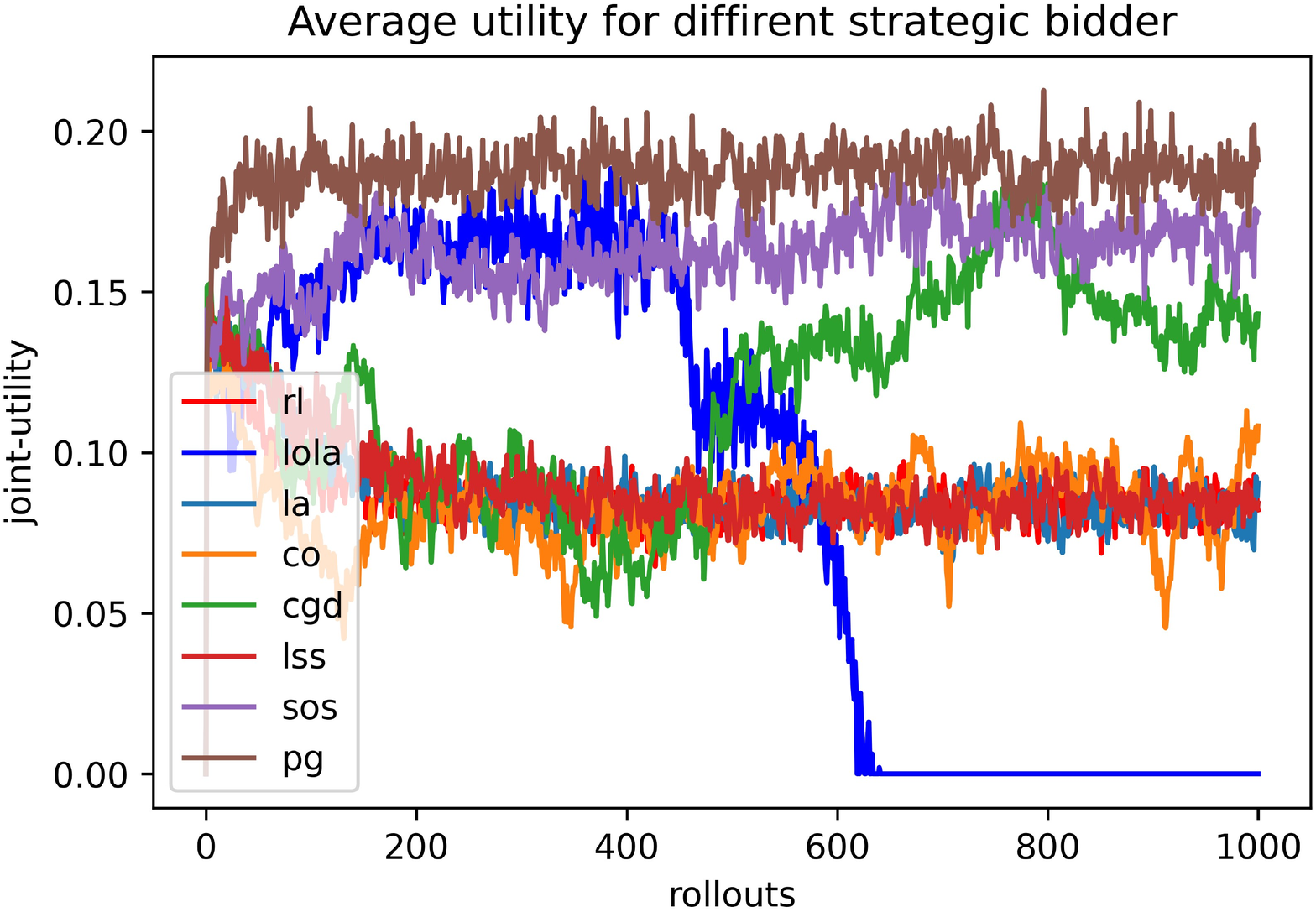}
  \vspace{-0.3cm}
  \caption{The average utility for the strategic bidders.}
  \label{fig6}
  \vspace{-0.3cm}
\end{figure}

\subsection{Experiment for PG Algorithm}

We compare the utility of agents with different strategy learning algorithms. The environment is the standard setting, and all bidders employ the linear shading strategy $B_i(v_i) = \alpha_i v_i$. In this environment, both bidders will adopt the same algorithm to learn the strategy. At the end of each auction round $t$, bidders and the seller will observe the strategies $(\alpha^t_i, \alpha^t_{-i}, \theta^t)$ and the corresponding rewards $(u^t_i, u^t_{-i}, r^t)$. Subsequently, the seller updates the mechanism in accordance with the Myerson Net, while the bidders adjust their strategies based on their respective learning algorithms. An equilibrium of the above system is $\alpha_1 = \alpha_2 = \frac{5}{14}$.

\balance

We compare the PG algorithm with opponent modeling and equilibrium solving algorithms (RL, LOLA \cite{no18}, LA \cite{no26}, CO \cite{no23}, CGD \cite{no24}, LSS \cite{no25}, SOS \cite{no19}) in the aforementioned environment. Figure 5 illustrates the evolution of agent strategy parameters during training. The results demonstrate that the majority of algorithms converge towards the truthful bidding strategy, aligning with the conclusion in Theorem 3.1. Algorithms that incorporate opponent predictions improve utility but display a degree of instability. Notably, the PG algorithm's training results closely approximate the Nash equilibrium strategy. The slight deviation from equilibrium could be attributed to the implementation of a sufficiently small reserve price. The average utility of various agent systems is shown in Figure 6, where the PG algorithm attains the highest average utility in the automated bidding task, consistent with Theorem 3.2.

\subsection{Experiments with Different Opponents and Environment Settings}

We conducted a series of experiments within diverse environments and with varying opponent strategies. Experiments in which other bidders use static strategies include: \ding{172} \textbf{Standard setting with a truthful bidding opponent.} In this scenario, another bidder consistently employs the truthful bidding strategy. \ding{173} \textbf{Standard setting with a Nash equilibrium opponent.} In this setup, another bidder adheres to the Nash equilibrium bidding strategy. \ding{174} \textbf{Asymmetric value distribution.} Here, the private value distribution of bidders exhibits asymmetry. \ding{175} \textbf{Dynamic value distribution.}  In this experiment, the value distribution function of the strategic bidder evolves over time. \ding{176} \textbf{Single-item three-bidders auction.} This experiment extends to a scenario with more than two bidders.

Experiments involving dynamic strategies employed by other bidders encompass: \ding{177} \textbf{LOLA algorithm opponent.} Here, another bidder updates its strategy using the LOLA algorithm. \ding{178} \textbf{SOS algorithm opponent.} In this case, another bidder adjusts its strategy using the SOS algorithm. In the two sets of experiments described above, the opponent bidder uses linear shading strategies.

We evaluate the effectiveness of the algorithms to improve the utility of the strategic bidder after 400 iterations. The results are presented in Table 1. The algorithms used for comparison (RL, LOLA, SOS) are experimented with both Bid Net and linear shading strategies, and the results in the table are taken from the better of the two. We see that the Bid Net trained with PG always maximizes the utility of the strategic bidder. More details of the setting and figures of the experiment are given in the supplementary material.


\section{Conclusion}

In this study, we have developed an automatic bidding approach for bidders in repeated auctions. We introduce the Bid Net as a representation of bidding strategies, and we have proposed the PG algorithm to train this network. We have shown the effectiveness of PG in learning optimal responses when faced with static opponents, as well as its convergence to induced equilibriums when all agents adopt PG simultaneously. Through a series of experiments, we have highlighted the superiority of Bid Net over the linear shading function and showcased the efficacy of the PG algorithm by comparing it with other opponent modeling algorithms. PG has proven to significantly enhance the utility of strategic bidders in varying environments and with diverse strategies employed by other agents. We hope that this work will contribute to more research on strategic bidders in auctions and automatic bidding.

\begin{acks}
This paper is supported by National Key R\&D Program of China (2021YFA1000403), and National Natural Science Foundation of China (Nos.11991022), and the DNL Cooperation Fund, CAS (DNL202023).
\end{acks}






\bibliographystyle{ACM-Reference-Format} 
\bibliography{bibtex}


\end{document}